\pdfoutput=1

\documentclass[aip,pop,preprint,amsmath,amssymb,superscriptaddress,floatfix]{revtex4-2}

\usepackage{graphicx}
\usepackage{booktabs}
\usepackage{multirow}
\usepackage{bm}
\usepackage[colorlinks=true,linkcolor=blue,citecolor=blue,urlcolor=blue]{hyperref}

\graphicspath{{figures/}}

\newcommand{\dv}{\mathrm{d}^3v}                %
\newcommand{\FMa}{F_{M\alpha}}                 %

\newcommand{\Npj}{N^{pj}_\alpha}               %
\newcommand{\Mlk}{M^{\ell k}_\alpha}           %
\newcommand{\Hh}{\hat{H}}                      %
\newcommand{\Lag}[2]{L_{#1}^{#2}}              %

\newcommand{\Cpar}{\mathsf{C}_\parallel}       %

\newcommand{\nustara}{\nu_{*\alpha}}
\newcommand{\nustari}{\nu_{*i}}          %
\newcommand{\EM}{\mathrm{EM}}
\newcommand{\ET}{\mathrm{ET}}

\newcommand{\jboot}{J_{BS}}            %

\newcommand{\pysfincs}{\texttt{pysfincs}}
\newcommand{\SFINCS}{\texttt{SFINCS}}

\begin{document}

\title{Local Neoclassical Calculations using a Hermite--Laguerre velocity-space basis with advanced collision operators}

\author{B. J. Frei}
\email[]{baptiste.frei@ipp.mpg.de}
\affiliation{Max-Planck-Institut f\"ur Plasmaphysik, Boltzmannstra\ss e 2,
85748 Garching, Germany}

\date{\today}

\begin{abstract}
We present radially local neoclassical (NC) calculations based on a Hermite--Laguerre (HL) velocity-space polynomial expansion combined with advanced linearized collision operators, including the exact Fokker--Planck--Landau (FPL) operator. The implementation is verified against the NC code \SFINCS{} using both a multi-species Lenard--Bernstein--Dougherty-like (LBD) operator and the FPL operator, the NC fluxes agreeing to within $3\%$ in the Pfirsch--Schl\"uter regime. The accuracy of the HL basis is then assessed by convergence studies against the Legendre--Laguerre (LG) basis. At an equal number of spectral coefficients, neither basis is uniformly more accurate. In the plateau regime, the LG basis is more accurate on the particle and heat fluxes, which are sensitive to the trapped--passing boundary, whereas the HL basis is more accurate on the bootstrap current, carried by the parallel flow. In the banana regime, both bases converge slowly. Finally, we compare the LBD and the original (OS) and improved Sugama (IS) operators against the FPL operator. The IS operator retrieves the NC fluxes to within $6\%$ over four decades of collisionality, whereas the LBD operator overestimates the ion heat flux by a factor of $2.6$ to $4.7$. The OS operator overestimates the ion heat flux by $4$ to $33\%$, while retrieving the particle flux and the bootstrap current to within $9\%$ and $18\%$, respectively. The convention adopted for the LBD collision frequency also changes the bootstrap current by up to two orders of magnitude. The present work extends the use of the HL basis, previously used in turbulent calculations, to NC physics and, to the knowledge of the author, these are the first NC calculations using the HL basis.
\end{abstract}

\maketitle

\section{Introduction}
\label{sec:intro}

Neoclassical (NC) transport is the collisional transport that survives in tokamaks \cite{wesson2011} and stellarators \cite{helander2014} once all fluctuations are removed, and it therefore sets the irreducible transport level of these devices \cite{hintonhazeltine1976,hirshmansigmar1981,helandersigmar2002}. It originates from the banana orbits of trapped particles, whose width, much larger than the Larmor radius, sets the step of the collisional random walk associated with small-angle deflections. The same trapped--passing collisional coupling also generates the bootstrap current and the parallel and poloidal flows of each species \cite{hintonhazeltine1976,sauter1999}.

The role of NC physics is particularly important in the pedestal of high-confinement (H-mode) plasmas, where the gradients are steepest and the collisionality varies strongly across
the pedestal. There, the radial electric field well is dominated by the main-ion diamagnetic contribution, the poloidal flow being close to
its NC value \cite{viezzer2013,viezzer2014}, and its shear suppresses the edge turbulence \cite{biglari1990,diamond1994} and is connected to the access to H-mode \cite{schirmer2006,cavedon2020,plank2023}. The bootstrap current drives the
peeling--ballooning modes that limit the pedestal \cite{snyder2009}, while the ion poloidal Larmor radius approaches the pedestal width \cite{kagan2008,landreman2012,trinczek2023}. In fact, in the H-mode pedestal, where turbulence is strongly suppressed, the ion heat transport is expected to approach its NC level \cite{viezzer2017}.

In these conditions, describing the self-consistent interactions between turbulent and NC transport becomes unavoidable, and full-$f$ gyrokinetic (GK) simulations provide a first-principles framework to model both \cite{ku2009,michels2021}. However, full-$f$ GK simulations resolving the interplay between turbulent and NC transport are computationally expensive. To alleviate this cost, spectral approaches in velocity space have received renewed attention \cite{mandell2018,frei2023,mandell2024,frei2025}. These investigations have focused on turbulent (linear and nonlinear) calculations, for which advanced linearized multi-species collision operators have been developed and implemented using the Hermite--Laguerre (HL) basis, including the original Sugama (OS) \cite{sugama2009}, the improved Sugama (IS) \cite{sugama2019,frei2022}, and the exact Fokker--Planck--Landau (FPL) collision operators \cite{jorge2019,frei2021}. We remark that the HL basis has also been extended to full-$f$ GK simulations in X-point geometry \cite{frei2025}. While these previous investigations have shown the computational efficiency of the HL basis and its ability to retrieve the turbulent transport, its ability to capture the NC physics remains to be addressed.

The goal of the present paper is therefore twofold. First, we assess the accuracy of the HL basis to describe NC transport and bootstrap current. Second, leveraging the HL advanced collision operators \cite{frei2021,frei2022}, we test the multi-species Lenard--Bernstein--Dougherty (LBD) collision operator \cite{lenard1958,dougherty1964}, used in full-$f$ GK codes \cite{ulbl2022,francisquez2022}, against the OS, IS and FPL operators. We solve for the steady-state solution of the drift-kinetic (DK) NC equation in the absence of a radial electric field. To test the convergence of the HL basis, we also solve the DK NC equation using the Legendre--Laguerre (LG) polynomial basis, the Legendre expansion of the pitch-angle dependence being widely used in NC solvers \cite{vanrij1989,belli2008,belli2012,landreman2014}. Indeed, the LG basis is a natural choice to describe the trapped--passing boundary, which
is a structure in the pitch angle (Legendre) at fixed energy (Laguerre). On the other hand, the HL basis is natural
for the parallel dynamics. The HL implementations of the LBD and FPL collision operators are verified against the \SFINCS{} code \cite{landreman2014}. First, we show that, at equal velocity-space resolution, the LG basis is the more accurate on the particle and heat fluxes and the HL one on the bootstrap current. Then, we show that the LBD collision operator overestimates the NC ion heat flux by a factor between $2.6$ and $4.7$ with respect to the FPL operator, whereas the IS operator yields the smallest deviation on the NC fluxes. The OS operator yields intermediate deviations on the NC fluxes, overestimating the NC ion heat flux by $4$ to $33\%$ and retrieving the NC particle flux to within $9\%$, while its bootstrap current, within $18\%$ of the FPL one over the whole collisionality range, is closer to the FPL prediction than the IS one at high collisionality. Finally, we show that the convention adopted for the collision frequency entering the LBD operator \cite{ulbl2022}, i.e., the symmetric ($\ET$) or the detailed-balance ($\EM$) form (Appendix~\ref{app:lbd}), yields large deviations in the NC fluxes and bootstrap current. Indeed, the $\ET$ form yields an electron--ion friction smaller than the physical one by nearly the mass ratio, such that the NC particle flux collapses, while the bootstrap current can exceed the FPL prediction by more than an order of magnitude. To our knowledge, the present work reports the first NC calculations based on the HL basis.

The rest of the paper is organized as follows. Sec.~\ref{sec:model} presents the
radially local NC problem and its two spectral representations, together with the
collision operators considered. Sec.~\ref{sec:sfincs} reports the benchmarks against
\SFINCS{} and the convergence studies of the two bases. Sec.~\ref{sec:collcomp}
compares the collision operators. Our conclusions follow. The linearized
multi-species LBD operator is derived in both bases in Appendix~\ref{app:lbd}, and the
diagnostics are given in Appendix~\ref{app:diag}.

\section{Spectral Neoclassical Model}
\label{sec:model}

The steady-state NC DK equation for the perturbed distribution function $f_{\alpha 1}$, such that $f_\alpha = F_{M\alpha}+ f_{\alpha 1}$ with $f_{\alpha 1} / F_{M\alpha} \sim \rho_{\theta\alpha}/L_\perp \ll 1$ (with $\rho_{\theta\alpha} = \rho_\alpha q / \varepsilon$ the poloidal Larmor radius, $\rho_\alpha$ the thermal Larmor radius, $q$ the safety factor, $\varepsilon$ the inverse aspect ratio and $L_\perp$ the equilibrium scale length), is given, in the radially local limit, by \cite{hintonhazeltine1976,hirshmansigmar1981,belli2008,landreman2014}

\begin{align} \label{eq:dke_cf}
 v_\parallel \nabla_\parallel f_{ \alpha 1}
- \frac{\mu}{m_\alpha} \nabla_{\parallel} B  \frac{\partial f_{ \alpha 1}}{\partial v_\parallel} - \sum_{\beta} C_{\alpha \beta} = - \bm{V}_{D \alpha}\cdot \nabla \psi \partial_\psi F_{M \alpha}(\psi),
\end{align}
where $\bm{V}_{D \alpha} \cdot \nabla \psi = \left[\bm{b} \times\left(v_\parallel^2 \bm{\kappa} + (\mu / m_\alpha)  \nabla B\right) /\Omega_\alpha\right] \cdot \nabla \psi  =  v_\parallel^2  V_{D \alpha}^\kappa+ (\mu B / m_\alpha) V_{D \alpha}^{\nabla B}$ is the radial magnetic drift, being
$V_{D \alpha}^{\kappa} = \bm b \times \bm \kappa \cdot \nabla \psi / \Omega_\alpha$ and
$V_{D \alpha}^{\nabla B} = \bm b \times \nabla \ln B \cdot \nabla \psi / \Omega_\alpha$ its curvature and
$\nabla B$ contributions, respectively. Here,
$\bm{\kappa}=\bm{b}\cdot\nabla \bm{b}$ is the magnetic curvature, $\mu = m_\alpha
v_\perp^2 / (2B)$ the magnetic moment and $\Omega_\alpha=q_\alpha B/m_\alpha$ the
gyrofrequency. In Eq.~\eqref{eq:dke_cf}, $\nabla_\parallel = \bm{b}\cdot \nabla $ is the parallel gradient along $\bm B$. The right-hand side of Eq.~\eqref{eq:dke_cf} contains the drive from the equilibrium density and temperature gradients, such that

\begin{align}
  \partial_\psi F_{M\alpha}(\psi)
   = \left[ \partial_\psi \ln n_\alpha(\psi)
       +\left(\frac{m_\alpha v^2}{2 T_\alpha(\psi)}-\frac32\right) \partial_\psi \ln T_\alpha (\psi)\right] F_{M\alpha}(\psi),
  \label{eq:drive_flux}
\end{align}
with $v^2 = v_\parallel^2 + v_\perp^2$. The left-hand side of Eq.~\eqref{eq:dke_cf} contains, in order, the parallel streaming, the mirror force, i.e., the parallel acceleration $\dot v_\parallel = -(\mu/m_\alpha)\,\bm{b}\cdot\nabla B$ that follows from the conservation of $\mu$ and is responsible for the magnetic trapping of particles, and the linearized collision operator, discussed in Sec.~\ref{sec:collisions}. We remark that the NC DK equation Eq.~\eqref{eq:dke_cf} can be obtained from the linearized GK model \cite{doerk2012,oberparleiter2015,frei2023}, when evaluated in the radially local, axisymmetric limit $k_x=k_y=0$ and retaining the radial advection of the equilibrium distribution function by the magnetic drift, neglected in the GK $\delta f$ model.

We now present two spectral approaches to solve Eq.~\eqref{eq:dke_cf}, based on an HL (Sec.~\ref{sec:hl}) and an LG (Sec.~\ref{sec:lg}) expansion, their equivalence being discussed in Sec.~\ref{sec:basis_transform}. Throughout, the velocity-space coordinates are normalized to the thermal speed
$v_{t\alpha}(\psi)=\sqrt{2T_\alpha(\psi)/m_\alpha}$. More precisely, we write
$\bm{c}_\alpha=\bm{v}/v_{t\alpha}$ for the normalized velocity and $c_\alpha=|\bm{c}_\alpha|$
for its magnitude, and we define the parallel and perpendicular projections of $\bm c_\alpha$
by $c_{\parallel \alpha} = \bm{b}\cdot\bm{c}_\alpha$ and $c_{\perp\alpha}^{2}=c_\alpha^{2}-c_{\parallel\alpha}^{2}$,
respectively, being $\xi= c_{\parallel \alpha}/c_\alpha$ the cosine of the pitch angle relative
to the local field direction $\bm{b} = \bm B / B$. The equilibrium is the normalized Maxwellian
$\FMa(\psi) = n_\alpha(\psi) / (\pi^{3/2}v_{t\alpha}^{3}(\psi)) e^{-c_\alpha^{2}}$. In both
approaches, we expand the perturbed distribution $f_{\alpha1}$ onto a set of orthonormal polynomial
basis in the velocity-space coordinates $(c_{\parallel \alpha}, c_{\perp \alpha}^2 )$ for HL and $(\xi, c_{\alpha}^2 )$ for LG. While the LG basis is widely used in NC calculations \cite{hirshmansigmar1981,belli2008,landreman2014}, the HL basis is often preferred for turbulence calculations \cite{beer1995,mandell2018,mandell2024,frei2022,frei2023}.

\subsection{Hermite--Laguerre spectral expansion}
\label{sec:hl}

The HL representation expands $f_{\alpha1}$ on a Hermite basis of the parallel velocity and a Laguerre basis of the perpendicular energy,\cite{frei2023,frei2025}
\begin{equation}
  f_{\alpha1} =  \sum_{pj}\FMa \Npj \Phi_{pj}(\bm c_\alpha),
  \qquad
  \Phi_{pj}(\bm{c}_\alpha) = \Hh_{p}\!\bigl(c_{\parallel \alpha}\bigr)\,L_j \bigl(c_{\perp \alpha}^2\bigr),
  \label{eq:hl_expand}
\end{equation}
summed over $0\le p\le P$, $0\le j\le J$. Here $\Hh_{p} = H_p / \sqrt{2^p p !}$ is the normalized
Hermite polynomial and $L_{j}$ the Laguerre polynomial \cite{abramowitz1972}, chosen such that
\begin{equation}
  \frac{1}{\pi^{3/2}} \int d \bm c_{ \alpha} e^{-c_\alpha^{2}}\,
    \Phi_{pj}(\bm{c}_\alpha)\,\Phi_{p'j'}(\bm{c}_\alpha)
    \;=\; \delta_{pp'}\delta_{jj'} ,
  \label{eq:hl_ortho}
\end{equation}
and the spectral moments follow by projection,
\begin{equation}
  \Npj = \frac{1}{n_\alpha} \int \dv\; \Phi_{pj}(\bm{c}_\alpha)\, f_{\alpha1} .
  \label{eq:hl_project}
\end{equation}
From Eq.~\eqref{eq:hl_project}, the lowest moments $N^{00}_\alpha$, $N^{10}_\alpha$,
$N^{20}_\alpha$ and $N^{01}_\alpha$ are the density, parallel flow, parallel and
perpendicular pressure perturbation moments of $f_{\alpha 1}$, respectively.

Projecting Eq.~\eqref{eq:dke_cf} onto $\FMa\Phi_{pj}(\bm c_\alpha)$ maps
each velocity-space operator onto a sparse coupling matrix, such that the steady-state
NC problem of Eq.~\eqref{eq:dke_cf} becomes the spectral moment hierarchy
\begin{equation}
  \mathcal{L}_{\alpha p'j'}^{pj}\, N^{p'j'}_{\alpha}
  \;+\; \mathcal{C}_{\alpha\beta p'j'}^{pj}\, N^{p'j'}_{\beta}
  \;+\; \mathcal{D}^{pj}_{\alpha} \;=\; 0 ,
  \label{eq:hier_hl}
\end{equation}
with summation over the repeated indices $(p'j')$ and over the repeated species label
$\beta$. In Eq.~\eqref{eq:hier_hl}, the operator $\mathcal{L}_{\alpha p'j'}^{pj}$ is
associated with the parallel streaming and the mirror force, and is given by
\begin{equation}
   \mathcal{L}_{\alpha p'j'}^{pj}
   = -\,v_{t\alpha}\Bigl[
     \nabla_\parallel \bigl(\Cpar\bigr)^{pj}_{p'j'}
      -( \nabla_\parallel \ln B)\,\bigl(\mathsf{M}\bigr)^{pj}_{p'j'}\Bigr].
  \label{eq:Lop_hl}
\end{equation}
In Eq.~\eqref{eq:Lop_hl}, the matrix $\Cpar$ arises from the parallel streaming and, using the recursive properties of the Hermite polynomials, its non-zero entries are

\begin{align}
  \bigl(\Cpar\bigr)^{pj}_{\;\;p+1,j} &= \sqrt{\tfrac{p+1}{2}}, &
  \bigl(\Cpar\bigr)^{pj}_{\;\;p-1,j} &= \sqrt{\tfrac{p}{2}}.
  \label{eq:Cparmatrix}
\end{align}

Similarly, the coupling matrix $\mathsf{M}$ arises from the mirror force, its non-zero
entries being
\begin{align}
  \bigl(\mathsf{M}\bigr)^{pj}_{\;\;p+1,j} &= \tfrac{1}{\sqrt2}(j{+}1)\sqrt{p{+}1}, &
  \bigl(\mathsf{M}\bigr)^{pj}_{\;\;p-1,j} &= -\tfrac{1}{\sqrt2}\,j\sqrt{p},
  \nonumber\\
  \bigl(\mathsf{M}\bigr)^{pj}_{\;\;p+1,j-1} &= -\tfrac{1}{\sqrt2}\,j\sqrt{p{+}1}, &
  \bigl(\mathsf{M}\bigr)^{pj}_{\;\;p-1,j+1} &= \tfrac{1}{\sqrt2}(j{+}1)\sqrt{p}.
  \label{eq:Mmatrix}
\end{align}
While the streaming matrix $\Cpar$ couples only
neighboring Hermite indices, the mirror-force matrix $\mathsf{M}$ couples the Hermite and
Laguerre indices jointly. In Eq.~\eqref{eq:hier_hl}, $\mathcal{C}_{\alpha\beta p'j'}^{pj}$ represents the multi-species linearized collision operators detailed in Sec.~\ref{sec:collisions}, being either a sparse or a full matrix depending on the operator. Finally, $\mathcal{D}^{pj}_{\alpha}$ is the projection of the gradient drive
appearing on the right-hand side of Eq.~\eqref{eq:dke_cf} onto the HL basis, i.e.,
\begin{equation}
  \mathcal{D}^{pj}_{\alpha}
  = -\frac{1}{n_\alpha} \int \dv\; \Phi_{pj}(\bm{c}_\alpha) \bm V_{D \alpha} \cdot \nabla \psi
    \partial_\psi F_{M \alpha}(\psi).
  \label{eq:drive_hl}
\end{equation}
Using the definition of $\bm V_{D \alpha}$, Eq.~\eqref{eq:drive_hl} is evaluated in closed form,
such that
\begin{equation}
  \mathcal{D}^{pj}_{\alpha}
  = - \sum_{s\, =\, \kappa,\, \nabla B}V_{D \alpha}^s \Bigl[
      \partial_\psi \ln n_\alpha\, \bigl(\mathsf{D}^{s}_{n}\bigr)^{pj}
    + \partial_\psi \ln T_\alpha\, \bigl(\mathsf{D}^{s}_{T}\bigr)^{pj} \Bigr],
  \label{eq:drive_hl_explicit}
\end{equation}
being $\mathsf{D}^{s}_{n}$ and $\mathsf{D}^{s}_{T}$ the density and temperature gradient
drive matrices of the curvature ($s=\kappa$) and $\nabla B$ ($s=\nabla B$) contributions, whose non-zero entries are
\begin{align}
  \bigl(\mathsf{D}^{\kappa}_{n}\bigr)^{00} &= 1, &
  \bigl(\mathsf{D}^{\kappa}_{n}\bigr)^{20} &= \sqrt2,
  \nonumber\\
  \bigl(\mathsf{D}^{\kappa}_{T}\bigr)^{00} &= 1, &
  \bigl(\mathsf{D}^{\kappa}_{T}\bigr)^{20} &= \tfrac{5}{\sqrt2},
  \nonumber\\
  \bigl(\mathsf{D}^{\kappa}_{T}\bigr)^{40} &= \sqrt6, &
  \bigl(\mathsf{D}^{\kappa}_{T}\bigr)^{01} &= -1,
  \nonumber\\
  \bigl(\mathsf{D}^{\kappa}_{T}\bigr)^{21} &= -\sqrt2,
  \label{eq:Dkappamatrix}
\end{align}
for the curvature contribution and
\begin{align}
  \bigl(\mathsf{D}^{\nabla B}_{n}\bigr)^{00} &= 1, &
  \bigl(\mathsf{D}^{\nabla B}_{n}\bigr)^{01} &= -1,
  \nonumber\\
  \bigl(\mathsf{D}^{\nabla B}_{T}\bigr)^{00} &= 1, &
  \bigl(\mathsf{D}^{\nabla B}_{T}\bigr)^{20} &= \tfrac{1}{\sqrt2},
  \nonumber\\
  \bigl(\mathsf{D}^{\nabla B}_{T}\bigr)^{01} &= -3, &
  \bigl(\mathsf{D}^{\nabla B}_{T}\bigr)^{21} &= -\tfrac{1}{\sqrt2},
  \nonumber\\
  \bigl(\mathsf{D}^{\nabla B}_{T}\bigr)^{02} &= 2,
  \label{eq:Dgradbmatrix}
\end{align}
for the $\nabla B$ one.
The linear system of equations given in Eq.~\eqref{eq:hier_hl} is closed by truncation,
such that $N^{pj}_\alpha=0$ for $p>P$ or $j>J$.

\subsection{Legendre--Laguerre spectral expansion}
\label{sec:lg}

The alternative expansion resolves the pitch angle directly, using Legendre polynomials
of $\xi$ and associated Laguerre polynomials of the energy $c_\alpha^{2}$ \cite{abramowitz1972},
\begin{equation}
  f_{\alpha1}
    =  \sum_{\ell k} \FMa \Mlk \Psi_{\ell k}(\bm{c}_\alpha),
  \qquad
  \Psi_{\ell k}(\bm{c}_\alpha) \;=\; c_{\ell k}\, c_\alpha^{\,\ell} P_{\ell}(\xi)\, \Lag{k}{\ell+1/2}\!\bigl(c_\alpha^{2}\bigr),
  \label{eq:lg_expand}
\end{equation}
summed over $0\le\ell\le L$, $0\le k\le K$. The prefactor
$c_\alpha^{\,\ell}P_{\ell}(\xi)$ is a
homogeneous polynomial of degree $\ell$ in $\bm{c}_\alpha$, and the constant $c_{\ell k}^{2} = \sqrt{\pi}\,(2\ell+1)\,k!/[2\,\Gamma(k+\ell+3/2)]$ enforces the orthonormality
\begin{equation}
  \frac{1}{\pi^{3/2}}\!\int d \bm c_\alpha e^{-c_\alpha^{2}}\,
    \Psi_{\ell k}(\bm{c}_\alpha)\,\Psi_{\ell'k'}(\bm{c}_\alpha)
    \;=\; \delta_{\ell\ell'}\delta_{kk'} .
  \label{eq:lg_ortho}
\end{equation}
The spectral moments follow by projection,
\begin{equation}
  \Mlk = \frac{1}{n_\alpha} \int \dv\; \Psi_{\ell k}(\bm{c}_\alpha)\, f_{\alpha1} .
  \label{eq:lg_project}
\end{equation}
From Eq.~\eqref{eq:lg_project}, the lowest moments $M^{00}_\alpha$, $M^{10}_\alpha$ and
$M^{01}_\alpha$ are the density, parallel flow and scalar temperature
perturbation moments of $f_{\alpha 1}$, respectively.

The projected hierarchy has the same index structure as
Eq.~\eqref{eq:hier_hl}, now acting on $\Mlk$,
\begin{equation}
  \mathcal{L}_{\alpha \ell' k'}^{\ell k}\, M^{\ell'k'}_{\alpha}
  \;+\; \mathcal{C}_{\alpha\beta \ell' k'}^{\ell k}\, M^{\ell'k'}_{\beta}
  \;+\; \mathcal{D}^{\ell k}_{\alpha} \;=\; 0 ,
  \label{eq:hier_lg}
\end{equation}
with the streaming--mirror operator carried by the two pitch-angle shift
operators $\mathcal{G}^{\pm}_{\ell}$ and their weights $\Sigma^{\pm}_{\ell}$,
\begin{equation}
  \mathcal{L}_{\alpha \ell' k'}^{\ell k}
    \;=\; -\,v_{t\alpha}\Bigl[
      \Sigma^{+}_{\ell}\,\mathcal{G}^{+}_{\ell}\,\bigl(\mathsf{A}^{+}\bigr)^{\ell k}_{\ell'k'}
      +\Sigma^{-}_{\ell}\,\mathcal{G}^{-}_{\ell}\,\bigl(\mathsf{A}^{-}\bigr)^{\ell k}_{\ell'k'}\Bigr],
  \label{eq:Lop_lg}
\end{equation}
being the weights and the parallel operators, both diagonal in $(\ell,k)$, given by
\begin{align}
  \Sigma^{+}_{\ell} &= (\ell{+}1)\,\sigma^{+}_\ell, &
  \Sigma^{-}_{\ell} &= \ell\,\sigma^{-}_\ell,
  \nonumber\\
  \mathcal{G}^{+}_{\ell} &= \nabla_\parallel-\tfrac12(\ell{+}2)(\nabla_\parallel \ln B), &
  \mathcal{G}^{-}_{\ell} &= \nabla_\parallel+\tfrac12(\ell{-}1)(\nabla_\parallel \ln B),
  \label{eq:lg_diag}
\end{align}
with
\begin{align}
  \sigma^{+}_\ell &= \frac{1}{\sqrt{(2\ell{+}1)(2\ell{+}3)}}, &
  \sigma^{-}_\ell &= \frac{1}{\sqrt{(2\ell{+}1)(2\ell{-}1)}}.
  \label{eq:sigmadef}
\end{align}
The energy couplings are generated by the shift matrices
\begin{align}
  \bigl(\mathsf{A}^{+}\bigr)^{\ell k}_{\;\;\ell+1,k} &= \sqrt{k{+}\ell{+}\tfrac32}, &
  \bigl(\mathsf{A}^{+}\bigr)^{\ell k}_{\;\;\ell+1,k-1} &= -\sqrt{k},
  \nonumber\\
  \bigl(\mathsf{A}^{-}\bigr)^{\ell k}_{\;\;\ell-1,k} &= \sqrt{k{+}\ell{+}\tfrac12}, &
  \bigl(\mathsf{A}^{-}\bigr)^{\ell k}_{\;\;\ell-1,k+1} &= -\sqrt{k{+}1},
  \label{eq:Amatrix}
\end{align}
We remark that, at variance with the HL basis, the
parallel streaming and the mirror force are carried by the same shift operators
$\mathcal{G}^{\pm}_{\ell}$, the Legendre index being shifted by one and the Laguerre index
by at most one. As in the HL case, $\mathcal{C}_{\alpha\beta \ell' k'}^{\ell k}$ represents the collision operators and $\mathcal{D}^{\ell k}_{\alpha}$ the projection of the gradient drive onto the LG basis, i.e.,
\begin{equation}
  \mathcal{D}^{\ell k}_{\alpha}
  = -\frac{1}{n_\alpha} \int \dv\; \Psi_{\ell k}(\bm{c}_\alpha) \bm V_{D \alpha} \cdot \nabla \psi
    \partial_\psi F_{M \alpha}(\psi).
  \label{eq:drive_lg}
\end{equation}
As in the HL case, Eq.~\eqref{eq:drive_lg} is evaluated in closed form,
\begin{equation}
  \mathcal{D}^{\ell k}_{\alpha}
  = - \sum_{s\, =\, \kappa,\, \nabla B}V_{D \alpha}^s \Bigl[
      \partial_\psi \ln n_\alpha\, \bigl(\mathsf{D}^{s}_{n}\bigr)^{\ell k}
    + \partial_\psi \ln T_\alpha\, \bigl(\mathsf{D}^{s}_{T}\bigr)^{\ell k} \Bigr],
  \label{eq:drive_lg_explicit}
\end{equation}
being $\mathsf{D}^{s}_{n}$ and $\mathsf{D}^{s}_{T}$ the LG counterparts of the drive
matrices of Eqs.~\eqref{eq:Dkappamatrix} and \eqref{eq:Dgradbmatrix}, whose non-zero
entries are
\begin{align}
  \bigl(\mathsf{D}^{\kappa}_{n}\bigr)^{00} &= 1, &
  \bigl(\mathsf{D}^{\kappa}_{n}\bigr)^{20} &= \tfrac{2}{\sqrt3},
  \nonumber\\
  \bigl(\mathsf{D}^{\kappa}_{n}\bigr)^{01} &= -\sqrt{\tfrac23}, &
  \bigl(\mathsf{D}^{\kappa}_{T}\bigr)^{00} &= 1,
  \nonumber\\
  \bigl(\mathsf{D}^{\kappa}_{T}\bigr)^{20} &= \tfrac{4}{\sqrt3}, &
  \bigl(\mathsf{D}^{\kappa}_{T}\bigr)^{01} &= -\tfrac{7}{2}\sqrt{\tfrac23},
  \nonumber\\
  \bigl(\mathsf{D}^{\kappa}_{T}\bigr)^{02} &= \sqrt{\tfrac{10}{3}}, &
  \bigl(\mathsf{D}^{\kappa}_{T}\bigr)^{21} &= -\sqrt{\tfrac{14}{3}},
  \label{eq:Dkappamatrix_lg}
\end{align}
for the curvature contribution and
\begin{align}
  \bigl(\mathsf{D}^{\nabla B}_{n}\bigr)^{00} &= 1, &
  \bigl(\mathsf{D}^{\nabla B}_{n}\bigr)^{20} &= -\tfrac{1}{\sqrt3},
  \nonumber\\
  \bigl(\mathsf{D}^{\nabla B}_{n}\bigr)^{01} &= -\sqrt{\tfrac23}, &
  \bigl(\mathsf{D}^{\nabla B}_{T}\bigr)^{00} &= 1,
  \nonumber\\
  \bigl(\mathsf{D}^{\nabla B}_{T}\bigr)^{20} &= -\tfrac{2}{\sqrt3}, &
  \bigl(\mathsf{D}^{\nabla B}_{T}\bigr)^{01} &= -\tfrac{7}{2}\sqrt{\tfrac23},
  \nonumber\\
  \bigl(\mathsf{D}^{\nabla B}_{T}\bigr)^{02} &= \sqrt{\tfrac{10}{3}}, &
  \bigl(\mathsf{D}^{\nabla B}_{T}\bigr)^{21} &= \tfrac12\sqrt{\tfrac{14}{3}},
  \label{eq:Dgradbmatrix_lg}
\end{align}
for the $\nabla B$ one.
The linear system of equations given in Eq.~\eqref{eq:hier_lg} is closed by truncation,
such that $M^{\ell k}_\alpha=0$ for $\ell>L$ or $k>K$.

\subsection{Basis Transformation} \label{sec:basis_transform}

While the HL and LG polynomial velocity-space bases consider expansions in the
$(c_{\parallel \alpha}, c_{\perp \alpha}^2)$ and $(\xi, c_{\alpha}^2)$ variables, respectively, an analytical equivalence can be derived between the two bases. In
particular, a basis transformation can be obtained \cite{jorge2019,frei2021,frei2022}, such that
\begin{equation}
  N_\alpha^{pj}
  = \sum_{\ell\, =\, 0}^{p + 2j}\ \sum_{k\, =\, 0}^{j + \lfloor p/2 \rfloor}
    T_{\ell k}^{pj}\, M_\alpha^{\ell k},
  \label{eq:basis_transform}
\end{equation}
being $T^{pj}_{\ell k}$ the analytical basis transformation coefficients, given by
\begin{equation}
  T^{pj}_{\ell k}
  = \frac{1}{\pi^{3/2}} \int d \bm c_{\alpha}\, e^{-c_\alpha^{2}}\,
    \Phi_{pj}(\bm{c}_\alpha)\, \Psi_{\ell k}(\bm{c}_\alpha) .
  \label{eq:basis_overlap}
\end{equation}
Closed analytical expressions of Eq.~\eqref{eq:basis_overlap} can be found in, e.g., Refs.~\onlinecite{jorge2019,frei2021,frei2022}. We remark that the sums in
Eq.~\eqref{eq:basis_transform} are finite. Indeed, $\Phi_{pj}$ and $\Psi_{\ell k}$ are
polynomials of degree $p + 2j$ and $\ell + 2k$ in $\bm{c}_\alpha$, such that $T^{pj}_{\ell k}$ vanishes unless
$\ell + 2k \le p + 2j$.

The inverse basis transformation, relating $M_\alpha^{\ell k}$ to $N_\alpha^{pj}$, is given by the transpose of $T^{pj}_{\ell k}$, i.e.,
$\left(T^{-1}\right)^{\ell k}_{pj} = T^{pj}_{\ell k}$, such that
\begin{equation}
  M_\alpha^{\ell k}
  = \sum_{p\, =\, 0}^{\ell + 2k}\ \sum_{j\, =\, 0}^{k + \lfloor \ell/2 \rfloor}
    T_{\ell k}^{pj}\, N_\alpha^{pj},
  \label{eq:basis_transform_inv}
\end{equation}
where the upper bounds follow from the same degree constraint as in
Eq.~\eqref{eq:basis_transform}. As an
example, Eqs.~\eqref{eq:basis_transform} and \eqref{eq:basis_transform_inv} yield
$N^{20}_\alpha = \sqrt{2/3}\, M^{20}_\alpha - M^{01}_\alpha / \sqrt3$ and
$M^{20}_\alpha = \sqrt{2/3}\, N^{20}_\alpha + N^{01}_\alpha / \sqrt3$, relating the
parallel pressure perturbation to the pitch-angle anisotropy of $f_{\alpha 1}$.

From Eq.~\eqref{eq:basis_transform}, the HL moments truncated at $(P, J)$ are retrieved \textit{exactly}
from the LG ones provided that $L \ge P + 2J$ and $K \ge J + \lfloor P/2 \rfloor$. Similarly,
Eq.~\eqref{eq:basis_transform_inv} retrieves the LG moments truncated at $(L, K)$ from
the HL ones provided that $P \ge L + 2K$ and $J \ge K + \lfloor L/2 \rfloor$.

These conditions are not fulfilled when Eqs.~\eqref{eq:hier_hl} and \eqref{eq:hier_lg} are truncated at comparable $(P,J)$ and $(L,K)$, such that neither truncated basis spans the polynomial space retained by the other and the analytical equivalence is lost at finite resolution. The number of moments required to reach a given accuracy is therefore a property of the basis rather than of the underlying NC problem, and is assessed in Sec.~\ref{subsec:convergencestudies} by comparing $(P, J)$ \textit{vs.} $(L, K)$.

\subsection{Multi-Species Linearized Collision Operators}
\label{sec:collisions}

We consider a hierarchy of linearized collision operators, namely the multi-species LBD operator~\cite{lenard1958,dougherty1964,francisquez2022,ulbl2022}, the multi-species OS~\cite{sugama2009} and IS~\cite{sugama2019} operators and the linearized FPL operator, the last three being referred to as advanced collision operators. These have been derived on the HL basis previously~\cite{frei2021,frei2022} and are expressed on the LG basis by the basis transformation of Sec.~\ref{sec:basis_transform}, while the linearized LBD operator is derived on both bases in Appendix~\ref{app:lbd}. While the advanced operators are dense, the LBD operator is sparse on both bases, which makes it well suited to the convergence studies and to the comparison with \SFINCS{}. We
remark that the LBD operator is also widely used in full-$f$ turbulence simulations
\cite{michels2021,ulbl2022,mandell2020,hakim2020,francisquez2020,francisquez2022}, while the accuracy of its NC predictions
has not been assessed.

The reference collision frequency is set by the parameter $\nu_c$ \cite{oberparleiter2015} and is expressed in units of $c_\mathrm{ref}/L_\mathrm{ref}$ (with $c_\mathrm{ref}=\sqrt{T_\mathrm{ref}/m_\mathrm{ref}}$). We define the like-species frequency $\nu_\alpha=\nu_{\alpha\alpha}$ by
\begin{equation}
  \nu_\alpha = \frac{8\sqrt2}{3\sqrt\pi}  \frac{\hat n_\alpha Z_\alpha^{4}}{\hat m_\alpha^{1/2}\hat T_\alpha^{3/2}} \nu_c,
  \label{eq:nu_anchor}
\end{equation}
being $Z_\alpha = q_\alpha/e$ the normalized charge, $\hat m_\alpha = m_\alpha/m_\mathrm{ref}$,
$\hat T_\alpha = T_\alpha/T_\mathrm{ref}$ and $\hat n_\alpha =
n_\alpha/n_\mathrm{ref}$. For the
advanced models, the normalized matrices already carry the mass and temperature dependence of the pair, through $\hat m_\alpha/\hat m_\beta$ and $\hat T_\alpha/\hat T_\beta$. Therefore, the frequency multiplying them carries only the charge and density of the field species, i.e.,
\begin{equation}
  \nu^{\mathrm{adv}}_{\alpha\beta} =
   \nu_\alpha \frac{\hat n_\beta Z_\beta^{2}}{\hat n_\alpha Z_\alpha^{2}}.
  \label{eq:nu_ab_fpl}
\end{equation}
For the LBD operator, in contrast, a single scalar per pair carries the whole
mass and temperature dependence of the collision,
\begin{equation}
  \nu^{\mathrm{LBD}}_{\alpha\beta} = 2^{3/2} \frac{8\sqrt2}{3\sqrt\pi}
   \frac{\hat n_\beta \sqrt{\hat m_\alpha \hat m_\beta}
         Z_\alpha^{2} Z_\beta^{2}}
        {\bigl(\hat m_\alpha \hat T_\beta +\hat m_\beta \hat T_\alpha\bigr)^{3/2}}  \nu_c,
  \label{eq:nu_ab}
\end{equation}
the factor $2^{3/2}$ making it retrieve Eq.~\eqref{eq:nu_anchor} at
$\alpha=\beta$. Two conventions, $\ET$ and $\EM$, differing by a mass factor, are available for the inter-species LBD frequency (Appendix~\ref{app:lbd}), Eq.~\eqref{eq:nu_ab} being the $\ET$ form.

We also define the normalized collisionality
\begin{equation}
  \nustara = \frac{\nu_\alpha\,q\,R_0}{\varepsilon^{3/2}\,v_{t\alpha}}
  = \frac{8}{3\sqrt\pi}\,\frac{q\,R_0}{\varepsilon^{3/2}}\,
    \frac{\hat n_\alpha Z_\alpha^{4}}{\hat T_\alpha^{2}}\,\nu_c ,
  \label{eq:nu_star}
\end{equation}
being $R_0$ the major radius and $\varepsilon=r_0/R_0$. The normalized collisionality defines the banana (BA) regime for
$\nustara \ll 1$, the plateau (PT) regime for
$1 \ll \nustara \ll \varepsilon^{-3/2}$ and the Pfirsch--Schl\"uter (PS) regime for
$\nustara \gg \varepsilon^{-3/2}$. We remark that, being $\varepsilon^{-3/2} \simeq 14.7$ for the parameters considered here, $\nustari = 10$ lies close to the transition between the PT and PS regimes, and is referred to as the PS regime in the following.

\section{Benchmarks and Convergence Studies}
\label{sec:sfincs}

Before assessing the convergence properties of the two spectral bases, we benchmark the spectral NC solver against the continuum NC code \SFINCS{} \cite{landreman2014}, which provides an independent reference. \SFINCS{} solves the same radially local, steady-state DK equation, Eq.~\eqref{eq:dke_cf}, with the radial electric field also set to zero. In velocity space, \SFINCS{} expands the
pitch-angle dependence on Legendre polynomials, similarly to the LG
basis of Sec.~\ref{sec:lg}, but represents the energy dependence on a grid in $x = v / v_{t \alpha}$. For collisions, \SFINCS{} provides the exact linearized multi-species
FPL operator \cite{landreman2014,hirshmansigmar1981}. For the present
work, we have developed \pysfincs{}, a Python equivalent of \SFINCS{} that also provides the linearized LBD operator of Appendix~\ref{app:lbd}, such that both the LBD and FPL operators can be benchmarked.

In all the simulations performed in this work, we use the circular equilibrium model of \citet{lapillonne2009} and, if not mentioned otherwise, we consider the General Atomics (GA) standard
case \cite{dimits2000,doerk2012}, which considers kinetic deuterium ions and electrons, with the mass ratio $m_i/m_e = 3672$, at equal
temperatures $T_i = T_e$ and with the normalized gradients $a/L_{n_\alpha}=1$ and
$a/L_{T_\alpha}=3$ for both species, being $a$ the minor radius. The reference flux surface is located at $r/a=0.5$
with an inverse aspect ratio of $\varepsilon
= r/R_0 = 1/6$ and the safety factor $q=2$. Finally, the normalized ion sound Larmor radius is $\rho_* = \rho_\mathrm{ref}/a =
10^{-3}$. To discretize the field line direction in the circular equilibrium, we use a uniform
grid with $N_z=24$ grid points over $-\pi\le z<\pi$, with periodic parallel boundary
conditions.

In this section, we first benchmark in Sec.~\ref{subsec:collbenchmark} the LBD and FPL collision operators implemented in the HL spectral basis against those of \SFINCS{}. We then study the convergence of the HL and LG bases as a function of collisionality in Sec.~\ref{subsec:convergencestudies}.

\subsection{Collision Operator Benchmarks against \texorpdfstring{\SFINCS{}}{SFINCS}} \label{subsec:collbenchmark}

We focus on the NC particle flux $\Gamma_i$, on the electron and ion heat fluxes, $Q_e$ and $Q_i$, and on the bootstrap current $\jboot$, defined in Appendix~\ref{app:diag}. The present comparison aims to show that the deviations between the two codes are set by the velocity-space resolution and not by a difference in the collision operators, such that \SFINCS{} can be used as the reference in Sec.~\ref{subsec:convergencestudies}.

If not specified otherwise, all the \SFINCS{} results are obtained with $N_\xi = 34$
Legendre modes and a generalized Gauss--Laguerre energy grid, with $N_x = 9$ for the
LBD and $N_x = 8$ for the FPL operator. We have verified that
the energy grid is converged, the fluxes varying by less than $0.05\%$ over
$6\le N_x\le14$ with the LBD and by less than $0.2\%$ between $N_x=8$ and $N_x=11$ with the FPL operator.

\begin{figure*}
  \includegraphics[width=0.8\textwidth]{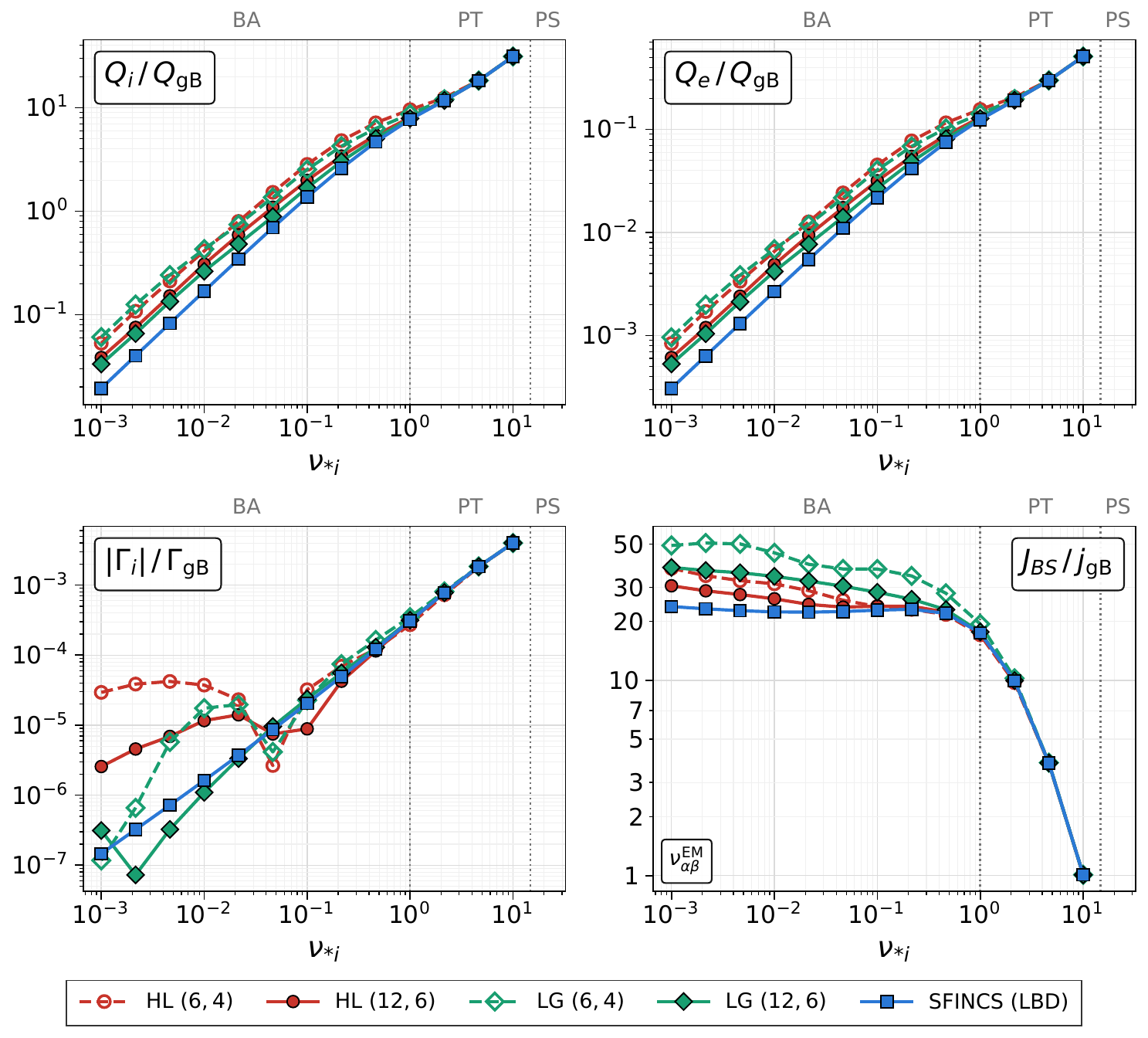}
  \caption{NC fluxes and bootstrap current normalized to gyro-Bohm (gB) units, $Q_i / Q_\mathrm{gB}$ (top left),
  $Q_e / Q_\mathrm{gB}$ (top right), $|\Gamma_i| / \Gamma_\mathrm{gB}$ (bottom left) and
  $\jboot / j_\mathrm{gB}$ (bottom right), as a function of $\nustari$, obtained with the LBD
  operator. The HL (red circles) and the LG (green diamonds) are shown with two
  velocity-space resolutions, $(6,4)$ (dashed lines, open symbols) and $(12,6)$ (solid
  lines, filled symbols), namely $(P,J)$ for HL and $(L,K)$ for LG, and \SFINCS{} by
  the blue squares. The fluxes use the LBD $\ET$ convention and $\jboot$ uses the $\EM$ convention (see Appendix~\ref{app:lbd}).}
  \label{fig:absolute_fluxes}
\end{figure*}

Figure~\ref{fig:absolute_fluxes} displays the NC fluxes, $Q_i$, $Q_e$, and $\Gamma_i$ computed with the LBD operator ($\ET$ convention) as a function of collisionality, from $\nustari = 10^{-3}$ (BA regime) to $\nustari = 10$ (PS regime). First, we observe that the HL, LG and \SFINCS{} results converge onto each other as the collisionality is increased, the three fluxes agreeing to $\lesssim 1.2 \%$ for $\nustari \gtrsim 5$. The deviations between the approaches are larger at lower collisionality, in the BA regime, being related to the appearance of velocity-space structures. Indeed, increasing the spectral resolution in both HL and LG from $(6,4)$ to $(12,6)$ improves the comparison with \SFINCS{}. Both observations demonstrate that the remaining deviations are related to the velocity-space resolution and not to a difference between the implementations. At $\nustari = 10$, the deviation with respect to \SFINCS{}, maximized over the three fluxes, is $0.08\%$ (HL) and $0.03\%$ (LG) already at $(6,4)$, and reaches $4\times10^{-3}\,\%$ at $(12,6)$ in both bases. Finally, the bootstrap current, obtained with the $\EM$ convention, is retrieved by both bases at high collisionality, with agreement below $5 \%$ for $\nustari \gtrsim 0.5$ at $(12,6)$. On the other hand, the deviation grows towards the BA regime, reaching $28\%$ for HL and $59\%$ for LG at $(12,6)$ and $\nustari = 10^{-3}$. Thus, $\jboot$ is sensitive to the velocity-space resolution at low collisionality, as investigated in Sec.~\ref{subsec:convergencestudies}.

\begin{figure*}
  \includegraphics[width=0.8\textwidth]{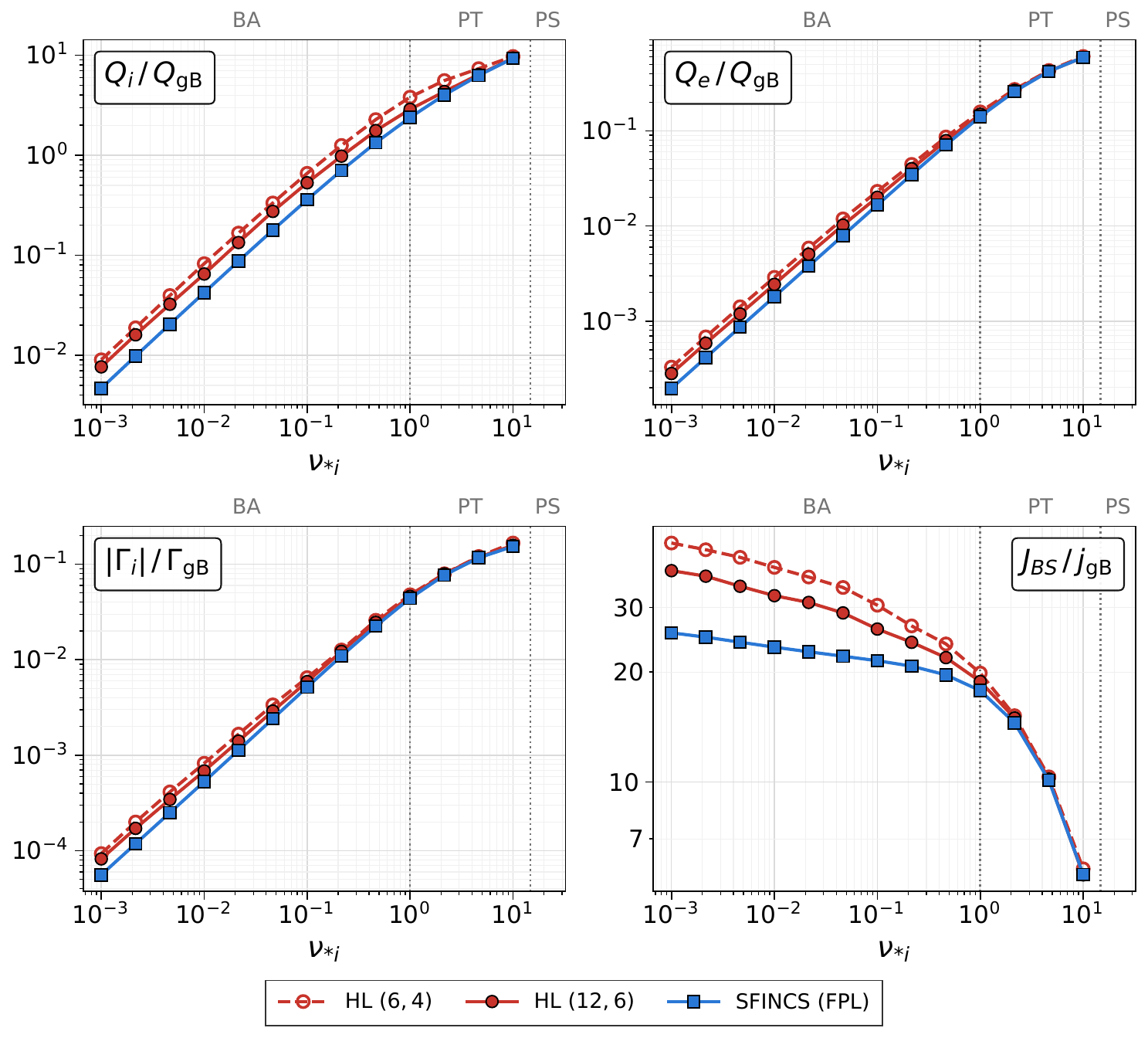}
  \caption{Same as Fig.~\ref{fig:absolute_fluxes}, with the FPL collision operator
  using HL and \SFINCS{}.}
  \label{fig:absolute_fluxes_fpl}
\end{figure*}

The benchmarks using the FPL operator are shown in Fig.~\ref{fig:absolute_fluxes_fpl} for the HL basis only, on which the FPL operator was originally developed \cite{frei2022,frei2023}. The observations made for the LBD operator also hold. More precisely, the deviation on $Q_i$ at the highest collisionality considered is $0.5\%$ at $(12,6)$ and reaches $0.2\%$ at $(16,8)$. In contrast, the deviation maximized over the three fluxes, as quoted above for the LBD operator, is set by $\Gamma_i$ and does not decrease with resolution, being $2.7\%$ at $(12,6)$ and $2.9\%$ at $(16,8)$. This indicates a residual difference of about $3\%$ on the particle flux between the two FPL implementations, which is not related to the HL velocity-space resolution. On the other hand, the deviations on the heat fluxes at low collisionality decrease with increasing $(P,J)$ and are therefore attributed to the velocity-space resolution. $\jboot$ behaves as in the LBD case, its deviation at $\nustari = 10^{-3}$ falling from $76\%$ at $(6,4)$ to $48\%$ at $(12,6)$, and reducing to $0.2\%$ at $\nustari = 10$. Overall, the two independent FPL implementations converge onto each other towards the PS regime, up to the residual difference on $\Gamma_i$. We remark that, contrary to the LBD operator, the HL expansion of the FPL operator involves the numerical evaluation of numerous analytical expressions subject to round-off errors \cite{frei2022}. Since the two codes implement the FPL operator independently, their agreement is itself a numerical verification of the HL FPL operator.

\subsection{Convergence Studies}
\label{subsec:convergencestudies}

We now assess the convergence properties of the HL basis, often considered in turbulence calculations, against those of the LG basis, conventionally preferred for NC calculations. Since their equivalence is lost at finite resolution (Sec.~\ref{sec:basis_transform}), we compare the two bases at similar $(P,J)$ and $(L,K)$ in different collisionality regimes, using the LBD operator throughout and taking \SFINCS{} as the reference.

 \begin{figure*}
  \includegraphics[width=0.9\textwidth]{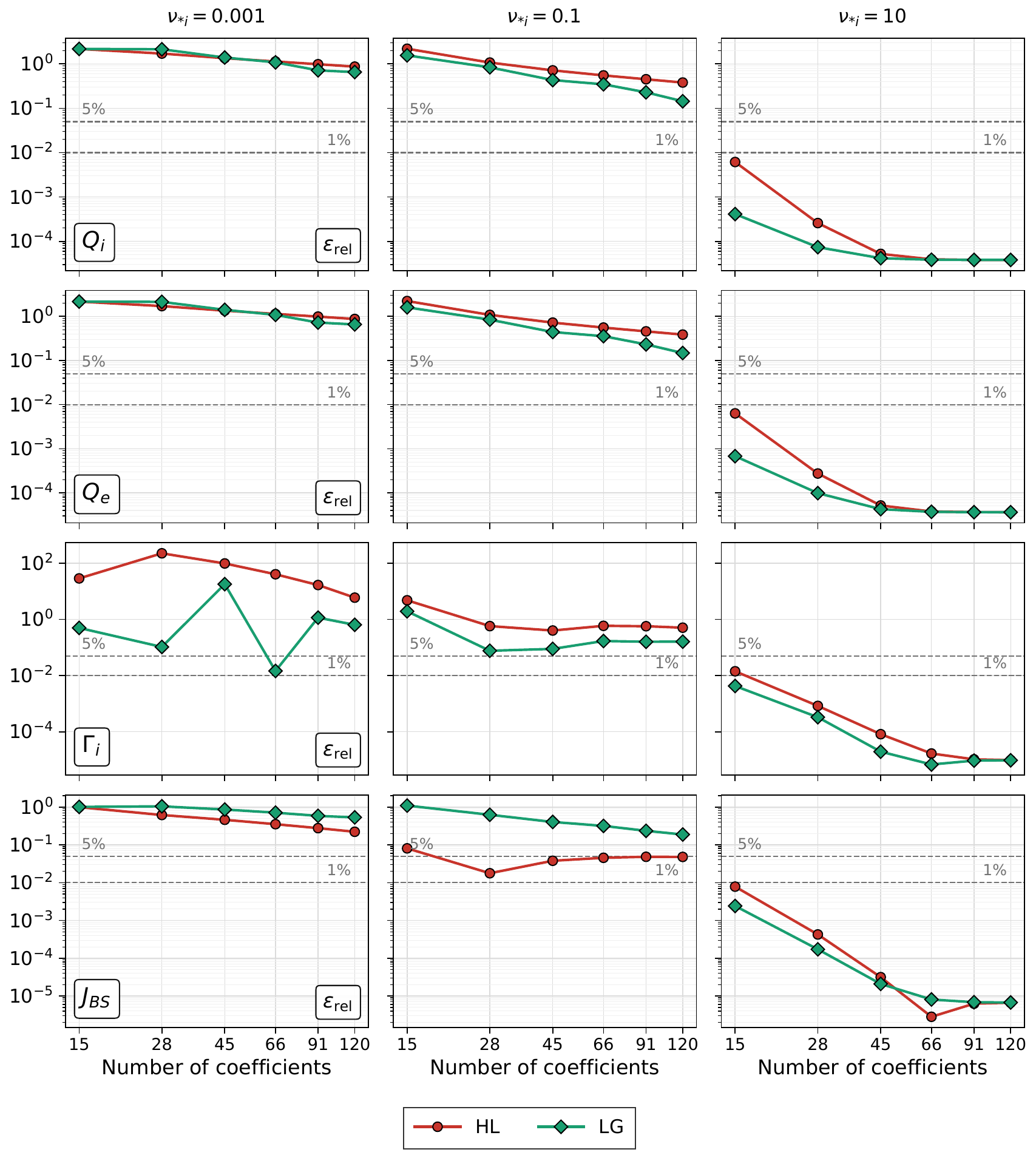}
  \caption{Relative error $\epsilon_{\rm rel}$ (with respect to \SFINCS{}) as a function
  of the number of coefficients, being $M=(P+1)(J+1)$ for HL and
  $M=(L+1)(K+1)$ for LG. The same spectral resolutions are used for both bases, i.e.,
  $(P,J) = (L,K) = (4,2)$, $(6,3)$, $(8,4)$, $(10,5)$, $(12,6)$ and $(14,7)$,
  corresponding to $M = 15$, $28$, $45$, $66$, $91$ and $120$, respectively.
  From top to bottom, $Q_i$, $Q_e$, $\Gamma_i$ and $\jboot$;
  from left to right, $\nustari = 10^{-3}$ (BA), $10^{-1}$ (PT) and $10$ (PS). HL is shown by the red
  circles and LG by the green diamonds, the dashed lines marking the $5\%$ and $1\%$
  levels of agreement with \SFINCS{}. The LBD operator is used throughout, with the $\ET$ convention for the fluxes and the $\EM$ convention for $\jboot$.}
  \label{fig:basisconvergence}
\end{figure*}

We measure the relative error $\epsilon_{\rm rel} = |X / X_\SFINCS{} - 1|$ of each NC diagnostic $X$ ($X = Q_i$, $Q_e$, $\Gamma_i$ and $\jboot$) as a function of the total number $M$ of retained coefficients, being $M = (P+1)(J+1)$ for HL and $M = (L+1)(K+1)$ for LG. Since the two bases are compared at the same $M$, the vertical distance between the two curves measures the accuracy gained by one basis over the other at an equal number of spectral coefficients.

Figure~\ref{fig:basisconvergence} displays $\epsilon_{\rm rel}$ for the four NC diagnostics in the BA, PT and PS collisionality regimes. First, at $\nustari = 10$ (PS) both bases converge steeply \cite{frei2021} and reach a floor near $\epsilon_{\rm rel} \simeq 4 \times 10^{-5}$ at $M \simeq 45$, this level being set by the \SFINCS{} discretization itself rather than by the spectral truncation.
Second, at $\nustari = 10^{-1}$ (PT), the LG
basis is the more accurate on the three fluxes, reaching $14\%$ on $Q_i$, $15\%$ on
$Q_e$ and $16\%$ on $\Gamma_i$ at $M = 120$, where HL remains at $38\%$, $39\%$ and
$50\%$ respectively. On the other hand, the HL basis is the more accurate on $\jboot$, which it
brings to $1.8\%$ already at $M = 28$ and keeps below $5\%$ over the whole ladder,
where LG remains above $18\%$. Third, at $\nustari = 10^{-3}$ (BA) neither
basis reaches the $5\%$ level even at $M = 120$, the error remaining of order unity on the
two heat fluxes, while on $\jboot$ the HL basis reaches $22\%$ against $53\%$ for the
LG one. We remark that the ordering of the two bases is therefore channel and regime dependent, the LG basis being the more accurate on the fluxes and the HL basis on the bootstrap current. Overall, the convergence is algebraic and slow in the BA and PT regimes, the error falling by less than a decade over the whole ladder.

\begin{figure*}
  \includegraphics[width=0.95\textwidth]{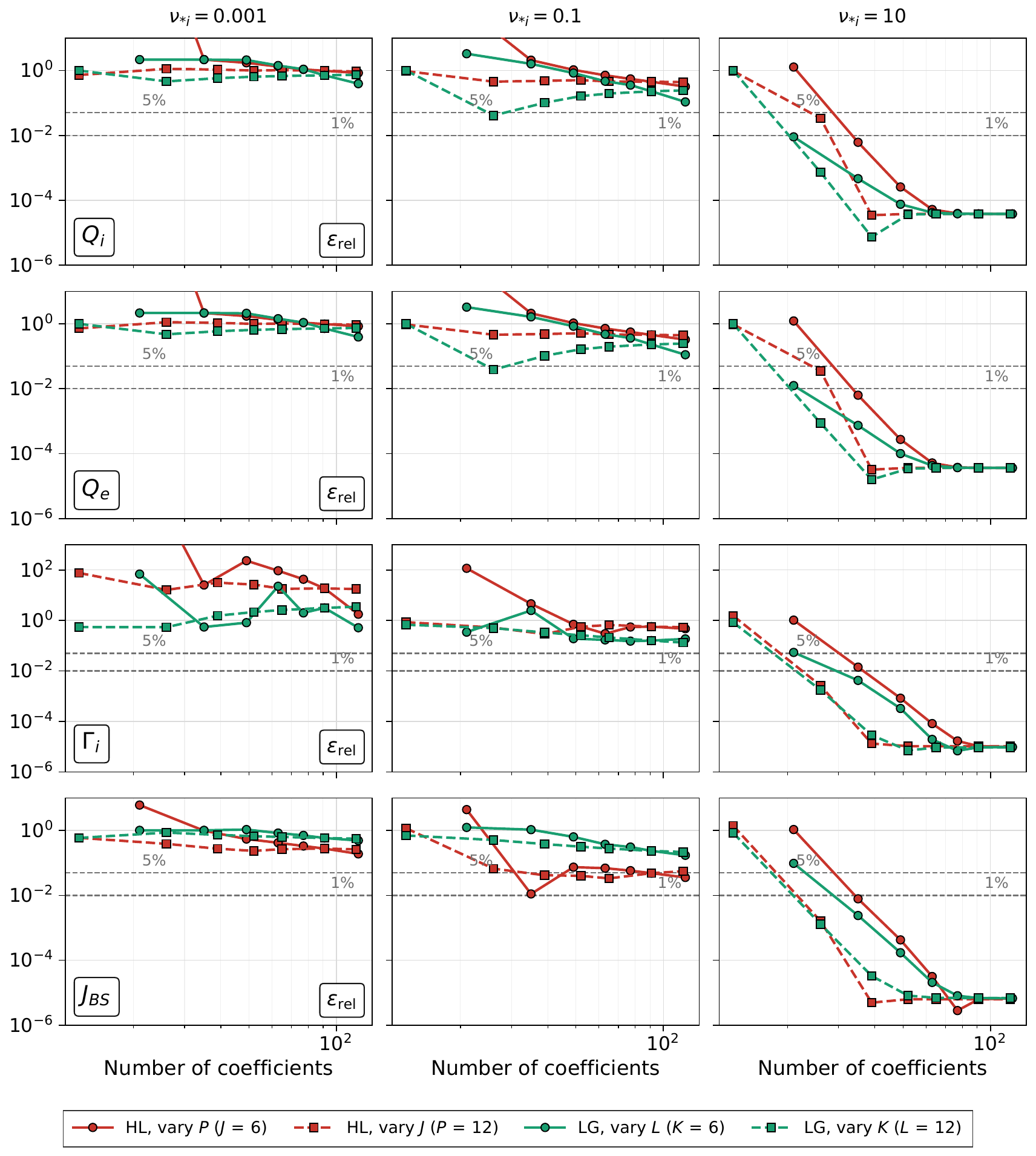}
  \caption{Relative error $\epsilon_{\rm rel}$ (with respect to \SFINCS{}) as a function of the number of coefficients $P$ ($L$) with $J$ ($K$) fixed, and vice versa. From top to bottom, $Q_i$,
  $Q_e$, $\Gamma_i$ and $\jboot$; from left to right, $\nustari = 10^{-3}$, $10^{-1}$
  and $10$. For HL (red), $P$ is varied at fixed $J = 6$ (circles, solid) and $J$ at
  fixed $P = 12$ (squares, dashed); for LG (green), $L$ at fixed $K = 6$ and $K$ at
  fixed $L = 12$. The dashed lines mark the $5\%$ and $1\%$ levels. The fluxes use the $\ET$ and
  $\jboot$ the $\EM$ convention.}
  \label{fig:directional}
\end{figure*}

To identify the velocity-space direction responsible for the convergence behavior of Fig.~\ref{fig:basisconvergence}, we increase a single truncation at a time, i.e., we fix $P$ ($L$) and vary $J$ ($K$), or vice versa, as shown in Fig.~\ref{fig:directional}. First, the error is controlled by the parallel order $P$ in the HL basis for all four NC diagnostics, the $J$ scan saturating at $J \simeq 2$, and by the pitch-angle order $L$ in the LG basis, the $K$ scan saturating at $K \simeq 2$ on $Q_i$. Second, and consistently with
Fig.~\ref{fig:basisconvergence}, the ordering of the two bases at
$\nustari \le 10^{-1}$ is set by the channel rather than by the direction. Indeed, on
the fluxes the LG error falls below the HL one, reaching
$11\%$ on $Q_i$ at $L = 16$ where HL remains at $32\%$, while the HL basis
is the more accurate on $\jboot$ by nearly an order of magnitude, being at $3.6\%$ at $P = 16$
where LG remains at $17\%$. We remark that this is consistent with the nature of the NC diagnostics. Indeed, the fluxes are carried by the trapped population, i.e., by a structure in the pitch angle resolved naturally by the Legendre expansion, whereas $\jboot$ is carried by the $\ell = 1$ harmonic of the parallel dynamics, directly associated with the Hermite expansion. The preferred basis is therefore set by the velocity direction carrying the diagnostic of interest.

\subsection{Collisionality Dependence of Velocity-Space Structures}

\begin{figure*}
  \includegraphics[width=\textwidth]{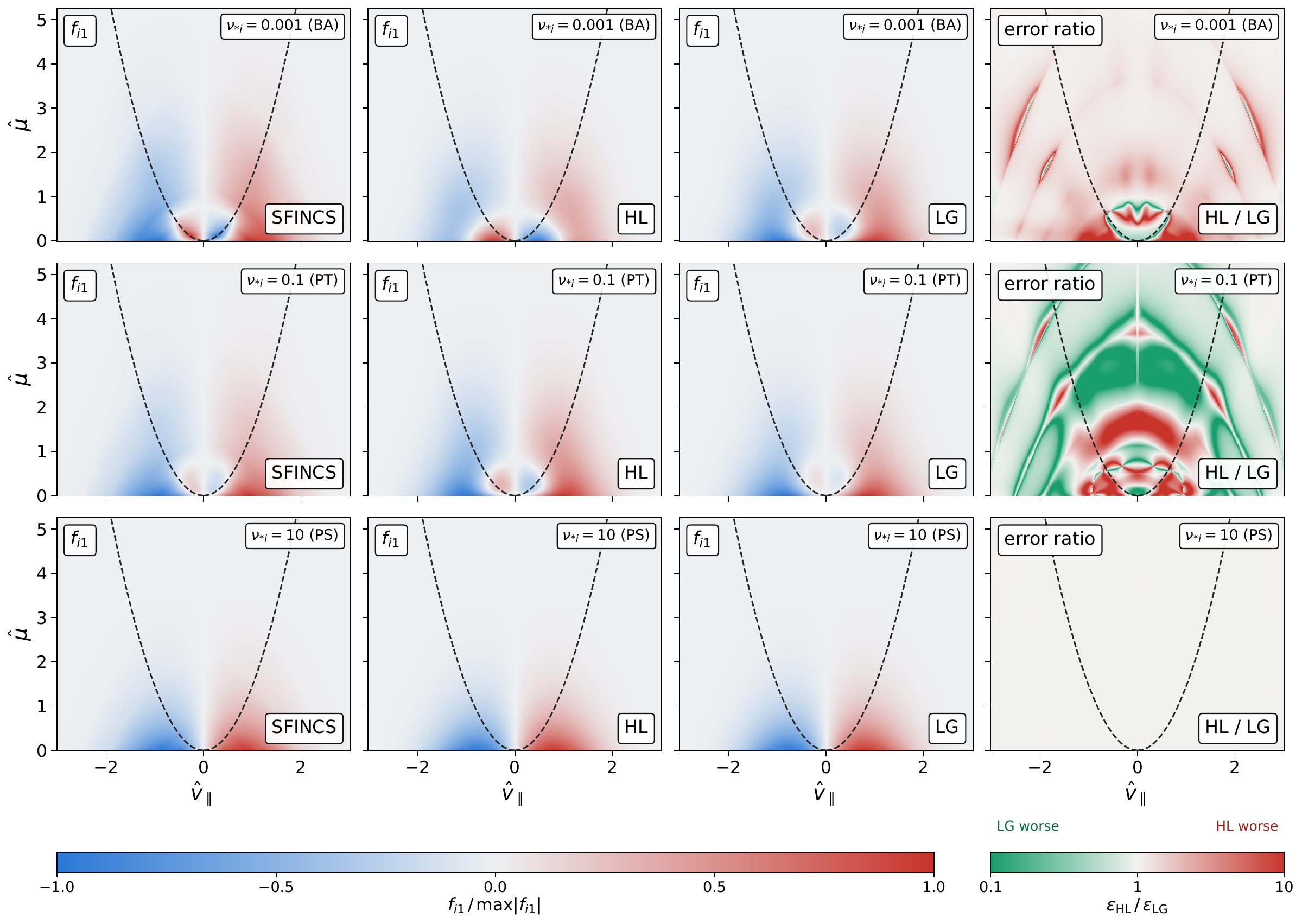}
  \caption{Ion distribution function $f_{i 1}$, (normalized to its maximum absolute
  value in each panel) in the $(\hat v_\parallel, \hat\mu)$ plane at the outboard midplane, being
  $\hat v_\parallel = c_{\parallel i}$ and $\hat\mu = \hat\mu_i = \mu B/T_{i0}$, obtained
  with \SFINCS{} (first column), with HL at $(P,J) = (16,8)$ (second) and with LG at
  $(L,K) = (16,8)$ (third). From top to
  bottom, $\nustari = 10^{-3}$ (BA), $10^{-1}$ (PT) and $10$ (PS). The rightmost column
  reports the ratio of the relative errors of the two bases (HL and LG) with respect to
  \SFINCS{}, $\epsilon_\mathrm{HL} / \epsilon_\mathrm{LG}$. The red (green) area denotes the region
  where the HL (LG) error dominates. The dashed line is the trapped--passing boundary.
  The LBD operator is used with the $\ET$ convention (see Appendix~\ref{app:lbd}).}
  \label{fig:distribution2d}
\end{figure*}

\begin{figure*}
  \includegraphics[width=\textwidth]{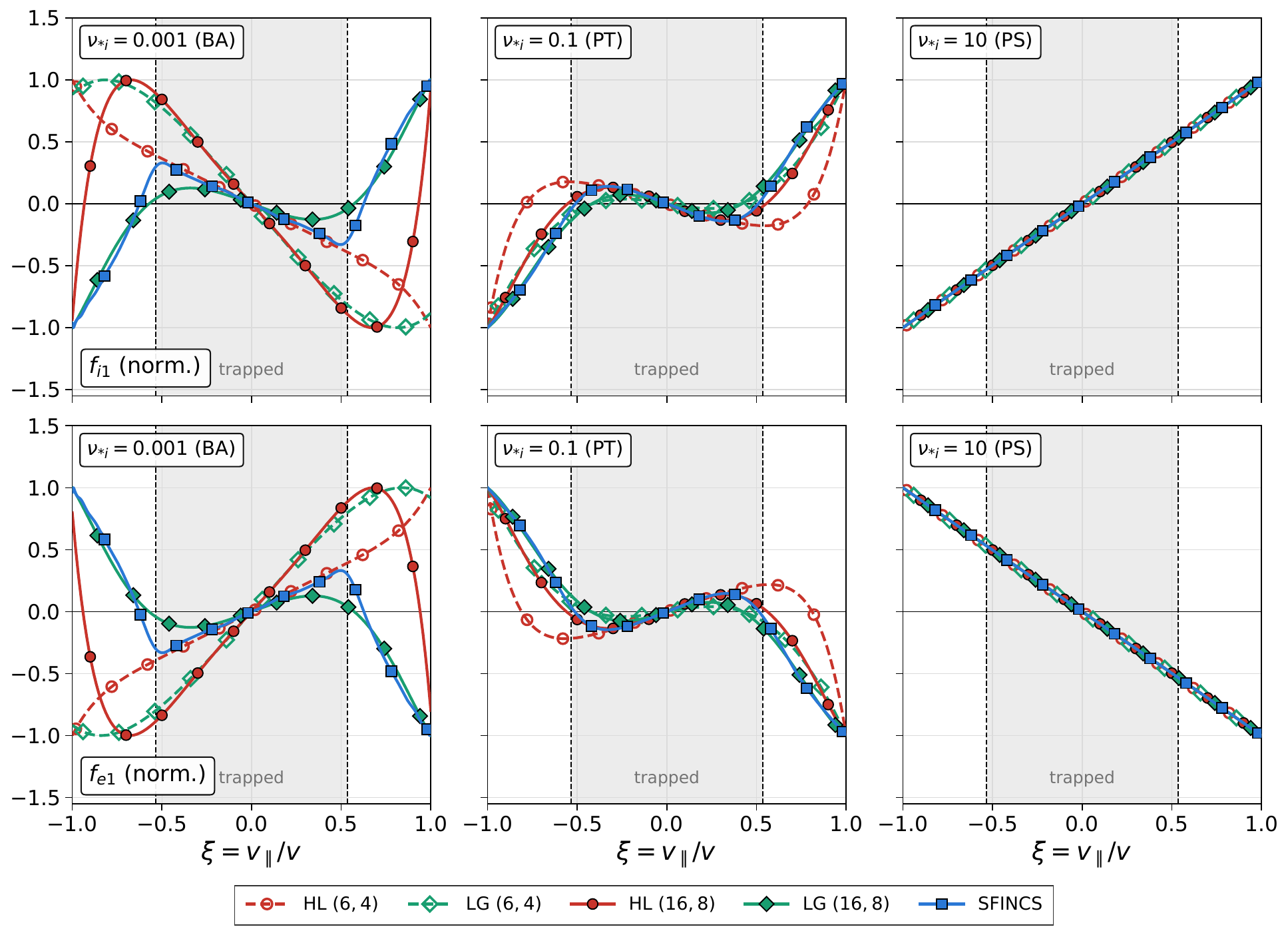}
  \caption{Perturbed distribution $f_{\alpha 1}$ as a function of the pitch angle $\xi$ at the
  outboard midplane and at the thermal speed, for the ions (top row) and the
  electrons (bottom row), at $\nustari = 10^{-3}$ (BA, left), $10^{-1}$ (PT, middle)
  and $10$ (PS, right). The HL (red circles) and LG (green diamonds) bases are shown
  at $(6,4)$ (dashed lines, open symbols) and $(16,8)$ (solid lines, filled symbols),
  and \SFINCS{} by the blue squares. The shaded band is the trapped region
  $|\xi| < \xi_c$, being $\xi_c = 0.5345$. Each curve is normalized
  to its own $\max|f_{\alpha 1}|$. The LBD operator is used with the $\ET$ convention (see
  Appendix~\ref{app:lbd}).}
  \label{fig:distributionpitch}
\end{figure*}

The origin of the collisionality dependence of the convergence properties is apparent in $f_{\alpha 1}$ itself, which we reconstruct from its spectral expansion in both bases and compare with \SFINCS{} on the same velocity-space grid. Figure~\ref{fig:distribution2d} displays the ion distribution function obtained with \SFINCS{}, HL and LG in the BA, PT and PS regimes, while Fig.~\ref{fig:distributionpitch} reports $f_{i 1}$ and $f_{e 1}$ as a function of $\xi$ across the trapped--passing boundary.

At $\nustari = 10$, $f_{i 1}$ is smooth and essentially proportional to $\xi$, i.e., carried by the single $\ell = 1$ harmonic, which both bases represent with few coefficients, yielding the fast convergence seen in
Fig.~\ref{fig:basisconvergence} in the PS regime.

As the collisionality is decreased in the PT and further in the BA regime, a fine structure develops at the trapped--passing boundary. This velocity-space structure can be observed in Fig.~\ref{fig:distribution2d} as a localized feature near
$\hat v_\parallel = 0$ and, in the BA regime, in Fig.~\ref{fig:distributionpitch} as local extrema near
$\xi = \pm\xi_c$, being $\xi_c = \sqrt{2\varepsilon/(1+\varepsilon)} = 0.5345$ the pitch angle at the trapped--passing boundary in the large aspect-ratio limit. The \SFINCS{} ion distribution has its extrema at $\xi \simeq \pm 0.50$ at $\nustari = 10^{-3}$, while they are found at $\xi \simeq \pm 0.33$ at $\nustari = 10^{-1}$, such that the localization at the trapped--passing boundary is a feature of the BA regime. Since this structure is non-analytic in the collisionless limit, its
spectral representation requires a number of coefficients that grows as the
collisionality is decreased, limiting the convergence of both spectral bases in the BA regime.

From the ratio of the two relative errors depicted in Fig.~\ref{fig:distribution2d}, the two bases fail differently near the trapped region, the same being observed on the pitch-angle structure of Fig.~\ref{fig:distributionpitch}. At $\nustari = 10^{-3}$, the LG result remains within $0.3$ of the \SFINCS{} one over the whole pitch-angle range, the curves being normalized to their own maximum, while the HL deviation reaches $1.4$, i.e., more than the peak amplitude. In addition, the HL basis places the extremum of $f_{i 1}$ at $\xi \simeq -0.68$ rather than at $\xi \simeq -0.50$ and overestimates its height by a factor of about three, such that the sign of $f_{i 1}$ is opposite over a large part of the passing region. On the other hand, the LG basis places the same extremum at $\xi \simeq -0.34$ and underestimates its height, its deviation from \SFINCS{} remaining roughly five times smaller than the HL one. The pitch-angle structure is therefore better resolved by LG, as
expected from its Legendre expansion, even at the collisionalities where HL is the
more accurate on the low-order moments of Fig.~\ref{fig:basisconvergence}. Indeed, as shown below, the NC diagnostics do not require a pointwise accurate $f_{i 1}$.

The bottom row of Fig.~\ref{fig:distributionpitch} reports the same
reconstruction for the electrons. In the BA regime at $\nustari = 10^{-3}$, neither basis accurately represents $f_{e 1}$ at the lowest resolution. Indeed, the LG $(6,4)$ reconstruction overshoots near the trapped--passing boundary, its slope across the trapped region exceeding the \SFINCS{} one, while the HL $(6,4)$ one reproduces that slope to $15\%$ but flattens the structure itself, carrying no extremum at $\xi \simeq \pm 0.50$. At $\nustari = 10^{-1}$, $f_{e 1}$ displays the same flattening across the trapped region as $f_{i 1}$, with opposite sign, such that the two species share the same pitch-angle structure. At $\nustari = 10$, the three representations are indistinguishable
for both species, $f_{\alpha 1}$ being then a straight line in $\xi$.

The pitch-angle content of $f_{\alpha 1}$ enters the NC diagnostics selectively, being
each of them a velocity-space integral of $f_{\alpha 1}$ (Appendix~\ref{app:diag}). Indeed, the radial magnetic drift weights $\Gamma_\alpha$
and $Q_\alpha$ by the parallel and perpendicular energies, i.e., by
$c_\alpha^{2}\xi^{2}$ and $c_\alpha^{2}(1-\xi^{2})$, which are even in $\xi$. More
precisely, Eq.~\eqref{eq:vdr_lg} expands these weights on $\Psi_{0k}$ and $\Psi_{2k}$
alone, such that the fluxes sample only the $\ell = 0$ and $\ell = 2$ harmonics of $f_{\alpha 1}$. On the
other hand, $\jboot$ is the parallel velocity moment, its weight being $c_\alpha \xi =
\Psi_{10}/\sqrt2$, such that it samples the $\ell = 1$ harmonic alone and is, in both
bases, associated with the single moment $N^{10}_\alpha$ or $M^{10}_\alpha$ carrying the parallel flow.

The two observations combine. Indeed, $f_{\alpha 1}$ is dominated by its $\ell = 1$ harmonic, being nearly proportional to $\xi$ in the PS regime, while its
$\ell = 0$ and $\ell = 2$ harmonics develop at the trapped--passing boundary as the collisionality decreases. Therefore, the flux channels are built from the part of $f_{\alpha 1}$ that the
spectral representation captures the least well, whereas $\jboot$ reads its dominant
harmonic directly. We remark that the deviations reported on $\jboot$ in
Sec.~\ref{subsec:convergencestudies} are therefore not a matter of representing the NC quantities, which a single moment carries exactly in both bases, but of how the
truncated hierarchy determines that moment. In fact, the flux weights
themselves are represented more efficiently in the LG basis, Eqs.~\eqref{eq:vdr_hl}
and \eqref{eq:vdr_lg} being exact for $L \ge 2$ and $K \ge 2$ while requiring
$P \ge 4$ and $J \ge 2$.

\section{Collision Operator Comparison}
\label{sec:collcomp}

We now use the HL formulation to compare the NC predictions of the LBD, OS and IS operators against the FPL operator, all the NC diagnostics being normalized to the FPL result obtained at the same $\nustari$ and resolution, $(P,J) = (12,6)$. We set $T_i/T_e = 1.5$, keeping the other GA parameters fixed, since the collision operator models are expected to differ precisely through the inter-species exchange terms, on which the temperature ratio acts \cite{frei2023}. The scan covers $10^{-3} \leq \nustari \leq 10$.

\begin{figure*}
  \includegraphics[width=0.8\textwidth]{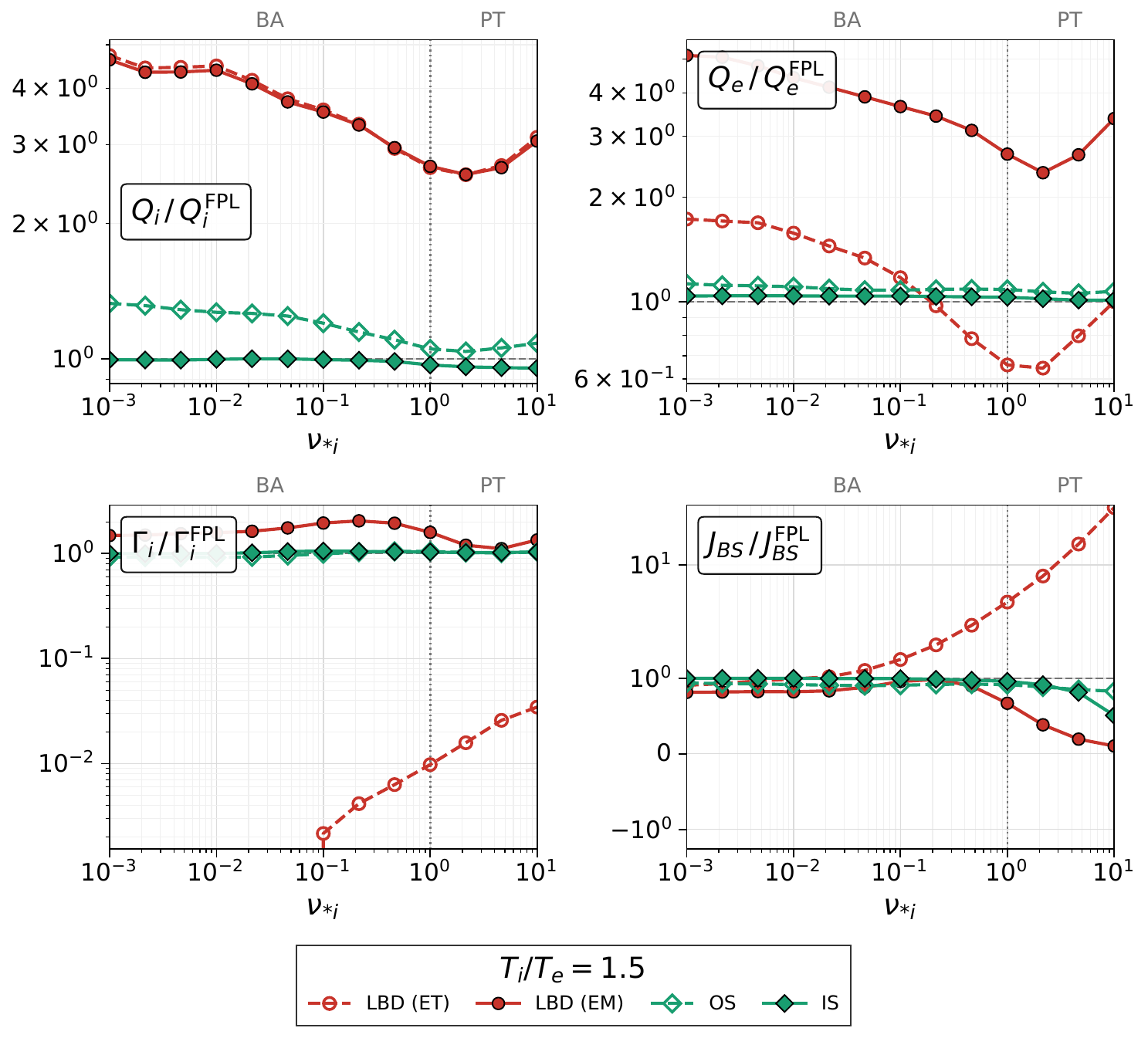}
  \caption{$Q_i$, $Q_e$, $\Gamma_i$ and $\jboot$ normalized by their FPL values as a
  function of $\nustari$, for LBD~($\ET$) (red dashed), LBD~($\EM$) (red solid), OS
  (green dashed) and IS (green solid). The horizontal line marks the FPL result. GA
  standard case with $T_i/T_e = 1.5$. The negative values of $\Gamma_i/\Gamma_i^\mathrm{FPL}$
  obtained with LBD~($\ET$) for $\nustari \lesssim 5\times10^{-2}$ are not displayed on the
  logarithmic axis.}
  \label{fig:collisionratiostau}
\end{figure*}

The collision operator comparison is displayed in Fig.~\ref{fig:collisionratiostau}. First, the IS operator retrieves
the FPL fluxes to a few percent, $Q_i$, $Q_e$ and $\Gamma_i$ remaining within $6\%$ of
the exact result over the whole scan. The IS bootstrap current agrees with the FPL one to better than $5\%$ for $\nustari \leq 1$, but departs from it at higher collisionality, reaching $0.81$ times the FPL current at $\nustari \simeq 4.6$ and $0.51$ times at $\nustari = 10$. Second, the OS operator produces a systematic and
comparatively modest deviation, overestimating $Q_i$ by $4$ to $33\%$ and $Q_e$ by less
than $13\%$, while $\Gamma_i$ is retrieved to better than $9\%$ and $\jboot$ to better than $18\%$ over the whole scan. We remark that, for $\nustari \gtrsim 4.6$, the OS bootstrap current is closer to the FPL one than the IS current, being $0.85$ times the FPL current at $\nustari \simeq 4.6$ and $0.83$ times at $\nustari = 10$. Third, the LBD operator
deviates from FPL by a much larger factor on the heat channels, its $Q_i$ exceeding the
exact result by a factor between $2.6$ and $4.7$ over the whole scan and in both $\ET$ and $\EM$ conventions.

The LBD comparison also isolates the role of the inter-species frequency convention (Appendix~\ref{app:lbd}). More precisely, $Q_i$ is almost
insensitive to the convention, its ratio to the FPL result decreasing from $4.7$ ($\ET$) and $4.6$ ($\EM$) at $\nustari = 10^{-3}$ to a minimum of $2.6$ at $\nustari \simeq 2$, before increasing again to $3.1$ at $\nustari = 10$ in both conventions. In contrast, $Q_e$ is strongly sensitive to the convention, being between $0.65$ and $1.7$ times the FPL result under $\ET$, while it is overestimated by a factor between $2.4$ and $5.1$ under $\EM$. The NC particle flux is also sensitive to the $\nu_{\alpha\beta}$ convention. Indeed, $\Gamma_i$ lies between $1.1$ and $2.0$ times the FPL result with $\EM$, whereas under $\ET$ it collapses, its magnitude remaining below $4\%$ of the FPL result over the
whole collisionality scan and its sign being reversed for $\nustari \lesssim 5\times10^{-2}$. The bootstrap current features qualitatively the same deviations. More precisely, the two conventions differ by more than two orders of magnitude at $\nustari = 10$,
where $\jboot$ falls to $0.10$ times the FPL current with $\EM$, while it is $32$
times larger than the FPL prediction with $\ET$. This sensitivity originates from the inter-species friction force. More precisely, the parallel momentum moment of the LBD operator of Eq.~\eqref{eq:lbd_fullf},
with the mixing flow of Eq.~\eqref{eq:lbd_mix}, yields the collisional friction force $R_{\alpha\beta}$ between
species $\alpha$ and $\beta$,
\begin{equation}
  R_{\alpha\beta} = m_\alpha n_\alpha \nu_{\alpha\beta}\,(u_{\alpha\beta}-u_\alpha)
  = m_\alpha n_\alpha\,\lambda_{\alpha\beta}\nu_{\alpha\beta}\,(u_\beta-u_\alpha).
  \label{eq:lbd_friction}
\end{equation}
The amplitude of $ R_{\alpha\beta}$ is set by the coefficient
$m_\alpha\lambda_{\alpha\beta}\nu_{\alpha\beta}$: with $\ET$,
$\nu^{\ET}_{\alpha\beta}=\nu_{\alpha\beta}$ and
$\lambda_{\alpha\beta}=m_\beta/(m_\alpha+m_\beta)$, while with $\EM$,
$\nu^{\EM}_{\alpha\beta}=\nu_{\alpha\beta}(m_\alpha+m_\beta)/(2m_\alpha)$ and
$\lambda_{\alpha\beta}=1/2$ (Appendix~\ref{app:lbd}), such that
\begin{equation}
  m_\alpha\lambda_{\alpha\beta}\nu^{\ET}_{\alpha\beta} = \mu_{\alpha\beta}\,\nu_{\alpha\beta}
  \qquad\text{and}\qquad
  m_\alpha\lambda_{\alpha\beta}\nu^{\EM}_{\alpha\beta} = \frac{m_\alpha+m_\beta}{4}\,\nu_{\alpha\beta},
  \label{eq:lbd_friction_coeff}
\end{equation}
being $\mu_{\alpha\beta}=m_\alpha m_\beta/(m_\alpha+m_\beta)$ the reduced mass. The
ratio of the $\EM$ to the $\ET$ friction coefficient is therefore
$(m_\alpha+m_\beta)^2/(4m_\alpha m_\beta)$, such that the $\EM$ friction is always
the stronger one. This ratio is equal to unity for equal masses but of the order of $m_i/m_e$ for electron--ion collisions.

We remark that it is the $\EM$ convention that retrieves the physical
scaling of the electron--ion friction force. Indeed, for $m_e \ll m_i$ and $T_e \sim T_i$,
Eq.~\eqref{eq:nu_ab} gives $\nu_{ei} \sim n_i\sqrt{m_e}/(m_i T_e^{3/2})$, such that
the $\EM$ coefficient, $(m_e+m_i) \nu_{ei} \sim
n_i\sqrt{m_e}/T_e^{3/2}$, which scales as $m_e$ times the Spitzer electron--ion collision
frequency, $\sim n_i/(\sqrt{m_e}\,T_e^{3/2})$ \cite{helandersigmar2002}. In contrast, the $\ET$ coefficient,
$\mu_{ei}\nu_{ei}\simeq m_e\nu_{ei}$, is smaller by the mass ratio, such that under the
$\ET$ convention the electron--ion friction is $\mathcal{O}(m_e/m_i)$ weaker than the physical one. Since the NC particle flux is
carried entirely by the inter-species friction, $\Gamma_\alpha$ collapses under $\ET$,
while $Q_i$, set by ion--ion collisions, is barely affected. Both conventions conserve
momentum exactly, such that the NC particle fluxes are ambipolar in the converged limit, and the
ambipolarity residual $|\Gamma_i - \Gamma_e|/\max_\alpha|\Gamma_\alpha|$ measures the
departure from ambipolarity due to the finite velocity-space resolution, relative to the flux amplitude.
At $\nustari = 10^{-3}$, this residual is $2\times10^{-3}$ for FPL and $1.1\times10^{-2}$ for
LBD--$\EM$. Under $\ET$, the residual also remains below $2\%$ for $\nustari \gtrsim 0.2$.
However, for $\nustari \lesssim 10^{-1}$, the friction-driven part of $\Gamma_\alpha$ is so
small under $\ET$ that the residual becomes comparable to the flux itself, reaching order
unity, such that the ambipolarity of $\Gamma_\alpha$ is no longer resolved. We note that
the same ordering of the operators and the same separation of the two LBD conventions
are obtained at $T_i = T_e$ (not shown), such that these conclusions are not an artifact of the temperature ratio.

Overall, the present study extends previous collision operator comparisons focused on microinstabilities \cite{frei2022,frei2023} to NC calculations, and confirms the good agreement between the IS and FPL operators on the fluxes. We remark that the $\ET$ convention is the one adopted in full-$f$ GK codes \cite{ulbl2022}, such that the deviations reported here can potentially affect these simulations.

\section{Conclusions}
\label{sec:conclusions}

In this work, we have presented radially local NC calculations performed, to our knowledge for the first
time, using the HL velocity-space basis, often preferred for turbulence calculations,
with advanced collision operators, including the exact linearized FPL operator. The
accuracy of the HL basis is assessed by comparing its NC predictions against the LG
basis, often preferred for NC calculations where the trapped and passing boundary is
naturally represented. Since the two representations are related by an exact basis
transformation, any deviation between them is attributable to the truncation alone. The two spectral approaches are successfully compared
with \SFINCS{} using the linearized multi-species LBD operator, derived here, and the FPL operator.

The present findings can be summarized as follows. First, an excellent agreement with \SFINCS{} is obtained with both the LBD and FPL operators. With the LBD operator, the NC fluxes agree to $\lesssim 1.2\%$ for $\nustari \gtrsim 5$. With the FPL operator, the heat fluxes agree to $0.5\%$ at $\nustari = 10$, while a residual difference of about $3\%$, independent of the HL resolution, remains on $\Gamma_i$. The bootstrap current is retrieved to better than $5\%$ for $\nustari \gtrsim 0.5$ with the LBD operator ($\EM$ convention) and to $0.2\%$ at $\nustari = 10$ with the FPL operator, while its deviation grows towards the BA regime,
reaching $28\%$ (HL) and $59\%$ (LG) at the same spectral resolution, $(P,J) = (L,K) = (12,6)$, with the LBD operator and $48\%$ (HL only) with the FPL operator.

Second, neither of the HL and LG approaches is uniformly more accurate with the same total number $M$ of spectral coefficients. Compared to \SFINCS{}, in the PT regime, the LG basis reaches a relative error of $14\%$ on $Q_i$ at $M = 120$ where
HL remains at $38\%$, while the HL basis yields an error below $5\%$ on $\jboot$ where LG remains above $18\%$. We remark that this ordering is set by the velocity direction carrying each diagnostic, the fluxes being carried by the trapped particles, i.e., by a structure in the pitch angle resolved naturally by the Legendre expansion, whereas $\jboot$ is carried by the $\ell = 1$ harmonic, directly associated with the Hermite expansion, as confirmed by the analysis of the velocity-space structures.

Third, we have compared different collision operator models against the FPL operator, namely the LBD, OS and IS operators. We have shown that the IS operator retrieves the NC fluxes within $6\%$ over the whole scan, $10^{-3} \leq \nustari \leq 10$. This justifies the use of IS, instead of FPL, for the NC fluxes when the computational cost of the latter is
prohibitive. We note, however, that the IS bootstrap current departs from the FPL one at high collisionality, being $0.51$ times the FPL current at $\nustari = 10$. On the other hand, the OS operator overestimates $Q_i$ by $4$ to $33\%$ and $Q_e$ by less than $13\%$, while $\Gamma_i$ is retrieved to better than $9\%$. In contrast, the LBD operator overestimates $Q_i$ by a factor between $2.6$ and $4.7$ in both $\ET$ and $\EM$ conventions.

Fourth, the choice of the inter-species collision frequency in the LBD operator \cite{ulbl2022} is not a free convention. Indeed, the $\ET$ convention reduces the electron--ion friction by a factor of the order of $m_i/m_e$ with respect to the $\EM$
convention, which retrieves the physical scaling of the electron--ion friction. As a consequence, using $\ET$, $\Gamma_i$ collapses below $4\%$ of the FPL result, with a sign reversal for $\nustari \lesssim 5\times10^{-2}$, and its ambipolarity
is no longer resolved for $\nustari \lesssim 10^{-1}$. Similarly, $\jboot$ exceeds the FPL current by a factor of $32$ under $\ET$ and is ten times smaller under $\EM$ at $\nustari = 10$. While $Q_i$ is almost insensitive to the convention, $Q_e$ is overestimated by a factor between $2.4$ and $5.1$ under $\EM$. We remark that the $\ET$ convention has been adopted in recent full-$f$ GK simulations of the edge and scrape-off layer \cite{michels2022,ulbl2023,frei2025,frei2025nf,ulbl2025,frei2026ppcf,frei2026prl}.

The present work therefore extends the previous investigations of the HL approach focusing on turbulence (see, e.g., \cite{frei2021,frei2022,frei2023}) to the NC problem. Overall, the HL basis performs as well as the LG basis in the PS regime, while in the PT regime the preferred basis depends on the NC quantity of interest. Leveraging the HL formulation of advanced collision operators, we have shown that the LBD operator yields order-unity (or larger) deviations in both the NC fluxes and the bootstrap current, a possible origin being its velocity-independent collision frequency, and that its inter-species frequency convention has a large impact on the NC particle flux and bootstrap current. In full-$f$ GK codes \cite{michels2021,ulbl2022,mandell2020,hakim2020,francisquez2020,francisquez2022}, where both turbulence and NC physics are retained, these deviations can potentially lead to large differences in the self-consistent interaction between turbulent fluxes and self-generated flows, in particular where turbulence is suppressed, such as in the H-mode pedestal, where the ion heat transport approaches its NC level \cite{viezzer2017}, and in optimized stellarators \cite{helander2014}. However, it remains unclear how these deviations translate into full-$f$ GK calculations, which is subject to future investigations.

\begin{acknowledgments}
The author would like to thank Philipp Ulbl, Andrew Sulimro, Akash Shukla, and Mattia Dicorato for helpful discussions about collision operators and neoclassical physics. This work has been carried out within the framework of the EUROfusion Consortium,
funded by the European Union via the Euratom Research and Training Programme (Grant
Agreement No 101052200 --- EUROfusion). Views and opinions expressed are however those
of the author(s) only and do not necessarily reflect those of the European Union or the
European Commission. Neither the European Union nor the European Commission can be held
responsible for them. The author used Claude Code (Anthropic) for
language editing, software development, and proofreading. The author
reviewed and edited all resulting changes, and takes full responsibility for the content of
this article.
\end{acknowledgments}

\section*{Author Declarations}
\subsection*{Conflict of Interest}
The author has no conflicts to disclose.

\section*{Data Availability}
The data that support the findings of this study are available from the corresponding
author upon reasonable request.

\appendix

\section{The linearized multi-species LBD collision operator}
\label{app:lbd}

We derive here the HL and LG spectral forms of the linearized multi-species LBD operator which reads, in the gyrocenter phase space $(v_\parallel,\mu)$ \cite{ulbl2022},
\begin{align}
  C_{\alpha\beta}(f_\alpha) = \nu_{\alpha\beta}\Biggl(
    &\frac{\partial}{\partial v_\parallel}\Bigl[
      \bigl(v_\parallel-u_{\alpha\beta}\bigr) f_\alpha
      +\frac{T_{\alpha\beta}}{m_\alpha}\frac{\partial f_\alpha}{\partial v_\parallel}\Bigr]
   + \frac{\partial}{\partial\mu}\Bigl[
      2\mu f_\alpha
      +\frac{2T_{\alpha\beta}}{B}\,\mu\,\frac{\partial f_\alpha}{\partial\mu}\Bigr]
  \Biggr).
  \label{eq:lbd_fullf}
\end{align}
The mixing flow $u_{\alpha\beta}$ and temperature $T_{\alpha\beta}$ are evaluated from
the moments of the distribution functions of both colliding species \cite{ulbl2022},
\begin{equation}
  u_{\alpha\beta} = (1-\lambda_{\alpha\beta})\,u_\alpha + \lambda_{\alpha\beta}\,u_\beta,
  \qquad
  T_{\alpha\beta} = (1-\kappa_{\alpha\beta})\,T_\alpha + \kappa_{\alpha\beta}\,T_\beta
     + \mathcal{O}\bigl(u^2\bigr),
  \label{eq:lbd_mix}
\end{equation}
being $u_\alpha$ and $T_\alpha$ the parallel flow and temperature of species
$\alpha$, and
\begin{equation}
  \lambda_{\alpha\beta} = \frac{m_\beta n_\beta\nu_{\beta\alpha}}
     {m_\alpha n_\alpha\nu_{\alpha\beta}+m_\beta n_\beta\nu_{\beta\alpha}},
  \qquad
  \kappa_{\alpha\beta} = \frac{n_\beta\nu_{\beta\alpha}}
     {n_\alpha\nu_{\alpha\beta}+n_\beta\nu_{\beta\alpha}}.
  \label{eq:lbd_weights}
\end{equation}
Eq.~\eqref{eq:lbd_mix} enforces the exact conservation of momentum and energy, with
$u_{\alpha\beta}=u_{\beta\alpha}$ and $T_{\alpha\beta}=T_{\beta\alpha}$. We remark that,
being $\nu_{\alpha\beta}\propto n_\beta$ and the temperature dependence of
Eq.~\eqref{eq:nu_ab} symmetric in $(\alpha,\beta)$, $\lambda_{\alpha\beta}$ and
$\kappa_{\alpha\beta}$ are constants, independent of the densities and temperatures,
with $\lambda_{\alpha\beta}+\lambda_{\beta\alpha}=\kappa_{\alpha\beta}+\kappa_{\beta\alpha}=1$.
The $\mathcal{O}(u^2)$ terms in Eq.~\eqref{eq:lbd_mix} are quadratic in the flows and
play no role at linear order.

We linearize Eq.~\eqref{eq:lbd_fullf} around a stationary Maxwellian background
$F_{M\alpha}$, of density $n_{\alpha0}$ and temperature $T_{\alpha0}$. At equilibrium,
$u_{\alpha\beta0}=0$ and $T_{\alpha\beta0}=\theta_{\alpha\beta}T_{\alpha0}$, being
$\theta_{\alpha\beta}$ the equilibrium mixing-temperature ratio which follows from
Eq.~\eqref{eq:lbd_mix},
\begin{equation}
  \theta_{\alpha\beta}
   =  \frac{1}{T_{\alpha 0 }}\left(\frac{n_{\alpha 0}\nu_{\alpha\beta}T_{\alpha 0 }+n_{\beta 0} \nu_{\beta\alpha}T_{\beta 0 }}
          {n_{\alpha 0}\nu_{\alpha\beta}+n_{\beta 0 }\nu_{\beta\alpha}} \right)
   = 1-\kappa_{\alpha\beta}+\kappa_{\alpha\beta}\,\frac{T_{\beta0}}{T_{\alpha0}}.
  \label{eq:lbd_Tmix}
\end{equation}
We remark that $\theta_{\alpha\beta}=1$ when $T_{\beta0} = T_{\alpha0}$, and in
particular that $\theta_{\alpha\alpha}=1$. Since $\lambda_{\alpha\beta}$ and
$\kappa_{\alpha\beta}$ are constants, the perturbed mixing moments follow directly
from Eq.~\eqref{eq:lbd_mix},
\begin{equation}
  u_{\alpha\beta1} = (1-\lambda_{\alpha\beta})\,u_{\alpha1}
                   + \lambda_{\alpha\beta}\,u_{\beta1},\qquad
  T_{\alpha\beta1} = (1-\kappa_{\alpha\beta})\,T_{\alpha1}
                   + \kappa_{\alpha\beta}\,T_{\beta1},
  \label{eq:lbd_mix1}
\end{equation}
and, therefore, depend on the perturbed flow and temperature, $u_{\alpha1}$ and
$T_{\alpha1}$, of both species. In addition, the perturbation of the collision
frequency, $\nu_{\alpha\beta1}$, acts on the equilibrium. Since
$\nu_{\alpha\beta}\propto n_\beta\,(m_\alpha T_\beta+m_\beta T_\alpha)^{-3/2}$,
$\nu_{\alpha\beta1}$ splits into a density part,
$\nu_{\alpha\beta}\,n_{\beta1}/n_{\beta0}$, which depends on the partner species alone,
and a temperature part, which is symmetric in $(\alpha,\beta)$. We retain the density
part, which is necessary for the conservation of energy at unequal equilibrium
temperatures, and we neglect the temperature part, which carries no net energy and
affects only the rate of the temperature equilibration. The linearized
LBD collision operator is decomposed into a \emph{test} ($\mathcal{C}^{T}_{\alpha\beta}$) and a \emph{field} ($\mathcal{C}^{F}_{\alpha\beta}$) part, such that
$C_{\alpha\beta}=\mathcal{C}^{T}_{\alpha\beta}+\mathcal{C}^{F}_{\alpha\beta}$. More
precisely, in the normalized velocity-space coordinates
$c_{\parallel\alpha}=v_\parallel/v_{t\alpha}$, $\hat\mu_\alpha=\mu B/T_{\alpha0}$, and
using Eq.~\eqref{eq:lbd_mix1}, the test and field components of the LBD collision
operator are given by
\begin{align}
  \mathcal{C}^{T}_{\alpha\beta} &= \nu_{\alpha\beta}\biggl[
      3f_{\alpha1}
      + c_{\parallel\alpha}\,\partial_{c_{\parallel\alpha}}f_{\alpha1}
      + \frac{ \theta_{\alpha\beta}}{2 } \partial^2_{c_{\parallel\alpha}}f_{\alpha1}
      + 2\hat\mu_\alpha\,\partial_{\hat\mu_\alpha}f_{\alpha1}
      + 2\theta_{\alpha\beta}\,\partial_{\hat\mu_\alpha}\bigl(\hat\mu_\alpha\,\partial_{\hat\mu_\alpha}f_{\alpha1}\bigr)
      \nonumber\\
    &\qquad\quad
      - (1-\lambda_{\alpha\beta})\,u_{\alpha 1}\,\partial_{v_\parallel}F_{M\alpha}
      + (1-\kappa_{\alpha\beta})\,T_{\alpha 1}\,\Theta_\alpha F_{M\alpha}
    \biggr],
  \label{eq:lbd_test_cont}\\
  \mathcal{C}^{F}_{\alpha\beta} &= \nu_{\alpha\beta}\Bigl[
      -\lambda_{\alpha\beta}\,u_{\beta 1}\,\partial_{v_\parallel}F_{M\alpha}
      + \Bigl(\kappa_{\alpha\beta}\,T_{\beta 1}
      + (\theta_{\alpha\beta}-1)\,T_{\alpha0}\,\frac{n_{\beta1}}{n_{\beta0}}\Bigr)
      \Theta_\alpha F_{M\alpha}
    \Bigr],
  \label{eq:lbd_field_cont}
\end{align}
respectively, being
$\Theta_\alpha g = \partial^2_{v_\parallel}g/m_\alpha
+ 2\,\partial_\mu(\mu\,\partial_\mu g)/B$, such that
$-u\,\partial_{v_\parallel}F_{M\alpha} = 2(u/v_{t\alpha})\,c_{\parallel\alpha}F_{M\alpha}$
and $T\,\Theta_\alpha F_{M\alpha} = (T/T_{\alpha0})\,
(2c^2_{\parallel\alpha}+2\hat\mu_\alpha-3)F_{M\alpha}$. In
Eqs.~\eqref{eq:lbd_test_cont} and \eqref{eq:lbd_field_cont},
$\nu_{\alpha\beta}\equiv\nu^{\mathrm{LBD}}_{\alpha\beta}$ is the collision frequency
given in Eq.~\eqref{eq:nu_ab}. The last two terms of Eq.~\eqref{eq:lbd_test_cont} originate from the dependence of the mixing moments on the
moments of $f_{\alpha1}$, Eq.~\eqref{eq:lbd_mix1}. Note that the last term of
Eq.~\eqref{eq:lbd_field_cont} is the density contribution from $\nu_{\alpha\beta1}$, which is proportional to
$(\theta_{\alpha\beta}-1)\,T_{\alpha0}\,\Theta_\alpha F_{M\alpha}$, and therefore
vanishes at equal equilibrium temperatures.

We remark that the differential part of $\mathcal{C}^{T}_{\alpha\beta}$, the first five terms of Eq.~\eqref{eq:lbd_test_cont}, can
be written as $\mathcal{C}^{T}_{\alpha\beta}=\nu_{\alpha\beta}\nabla_{\bm{c}_\alpha}\!\cdot\!
(\bm{c}_\alpha f_{\alpha1}+ \theta_{\alpha\beta} / 2 \nabla_{\bm{c}_\alpha}f_{\alpha1})$. In the case $\theta_{\alpha\beta}=1$, $\mathcal{C}^{T}_{\alpha\beta}$ admits both the Hermite--Laguerre \emph{and} the Legendre--Laguerre
basis functions as eigenfunctions, with the same eigenvalue $-(p+2j)=-(\ell+2k)$.

The perturbed density, flow and temperature entering Eqs.~\eqref{eq:lbd_test_cont}
and \eqref{eq:lbd_field_cont} are expressed in terms of the HL spectral moments as
\begin{equation}
  \frac{n_{\alpha1}}{n_{\alpha0}} = N^{00}_\alpha,
  \qquad
  \frac{u_{\alpha1}}{v_{t\alpha}} = \frac{N^{10}_\alpha}{\sqrt2},
  \qquad
  \frac{T_{\alpha1}}{T_{\alpha0}} = \frac13\Bigl(\sqrt2\,N^{20}_\alpha
     - 2N^{01}_\alpha\Bigr),
  \label{eq:lbd_uT}
\end{equation}
and similarly for species $\beta$, with $u_{\beta1}/v_{t\alpha} =
(v_{t\beta}/v_{t\alpha})\,N^{10}_\beta/\sqrt2$. Using
$-u\,\partial_{v_\parallel}F_{M\alpha} \mapsto \sqrt2\,(u/v_{t\alpha})\,\delta^p_1\delta^j_0$
and $T\,\Theta_\alpha F_{M\alpha} \mapsto (T/T_{\alpha0})\,(\sqrt2\,\delta^p_2\delta^j_0 -
2\,\delta^p_0\delta^j_1)$, the HL projection of the linearized LBD collision operator
yields
\begin{multline}
  \mathcal{C}^{Tpj}_{\alpha\beta} = \nu_{\alpha\beta}\Bigl[
    -(p+2j)\,N^{pj}_\alpha
    + \bigl(\theta_{\alpha\beta}-1\bigr)\Bigl(\sqrt{p(p-1)}\,N^{p-2,j}_\alpha
      - 2j\,N^{p,j-1}_\alpha\Bigr)
    + (1-\lambda_{\alpha\beta})\,N^{10}_\alpha\,\delta^p_1\delta^j_0 \\
    + \frac{1-\kappa_{\alpha\beta}}{3}\Bigl(\sqrt2\,N^{20}_\alpha - 2N^{01}_\alpha\Bigr)
      \Bigl(\sqrt2\,\delta^p_2\delta^j_0 - 2\,\delta^p_0\delta^j_1\Bigr)\Bigr],
  \label{eq:lbd_test_hl}
\end{multline}
for the test part and
\begin{multline}
  \mathcal{C}^{Fpj}_{\alpha\beta} = \nu_{\alpha\beta}\biggl\{
    \lambda_{\alpha\beta}\sqrt{\frac{m_\alpha T_{\beta0}}{m_\beta T_{\alpha0}}}\,
      N^{10}_\beta\,\delta^p_1\delta^j_0 \\
    + \biggl[\frac{\kappa_{\alpha\beta}}{3}\,\frac{T_{\beta0}}{T_{\alpha0}}
      \Bigl(\sqrt2\,N^{20}_\beta - 2N^{01}_\beta\Bigr)
      + (\theta_{\alpha\beta}-1)\,N^{00}_\beta\biggr]
      \Bigl(\sqrt2\,\delta^p_2\delta^j_0 - 2\,\delta^p_0\delta^j_1\Bigr)\biggr\},
  \label{eq:lbd_field_hl}
\end{multline}
for the field part. The self-restoring terms of Eq.~\eqref{eq:lbd_test_hl} act on the
momentum row $(1,0)$ and on the energy rows $(2,0)$ and $(0,1)$ only. For like-species
collisions, one has $\lambda_{\alpha\alpha}=\kappa_{\alpha\alpha}=\tfrac12$ and
$\theta_{\alpha\alpha}=1$, such that the self-restoring and field terms add up to
$N^{10}_\alpha$ and to $\frac13(\sqrt2\,N^{20}_\alpha - 2N^{01}_\alpha)$ with unit
coefficient. The $(1,0)$ row and the combination $\sqrt2\,\mathcal{C}^{20}_{\alpha\alpha}
- 2\,\mathcal{C}^{01}_{\alpha\alpha}$ then vanish identically, i.e., the like-species
operator conserves momentum and energy exactly.

On the LG basis, the same continuum operator,
Eqs.~\eqref{eq:lbd_test_cont}--\eqref{eq:lbd_field_cont}, is projected by using
$\nabla^2_{\bm{c}_\alpha}\Psi_{\ell k}
=-4(k+\ell+\tfrac12)(c_{\ell k}/c_{\ell,k-1})\Psi_{\ell,k-1}$ for the differential part.
The remaining terms are projected by noticing that $\Psi_{10}=\sqrt2\,c_{\parallel\alpha}$
and $\Psi_{01}=\sqrt{2/3}\,(3/2-c_\alpha^2)$, such that
$-u\,\partial_{v_\parallel}F_{M\alpha} = \sqrt2\,(u/v_{t\alpha})\,\Psi_{10}F_{M\alpha}$
and $T\,\Theta_\alpha F_{M\alpha} = -\sqrt6\,(T/T_{\alpha0})\,\Psi_{01}F_{M\alpha}$, and
that, from Eq.~\eqref{eq:lg_project}, $n_{\alpha1}/n_{\alpha0} = M^{00}_\alpha$,
$u_{\alpha1}/v_{t\alpha} = M^{10}_\alpha/\sqrt2$ and
$T_{\alpha1}/T_{\alpha0} = -\sqrt{2/3}\,M^{01}_\alpha$. The LG spectral expansion of the
LBD collision operator then reads
\begin{align}
  \mathcal{C}^{T\ell k}_{\alpha\beta} &= \nu_{\alpha\beta}\Bigl[-(\ell+2k)\,M^{\ell k}_\alpha
      + (\theta_{\alpha\beta}-1)\,g_{\ell k}\,M^{\ell,k-1}_\alpha
      \nonumber\\
    &\qquad\quad
      + (1-\lambda_{\alpha\beta})\,M^{10}_\alpha\,\delta_{\ell1}\delta_{k0}
      + 2(1-\kappa_{\alpha\beta})\,M^{01}_\alpha\,\delta_{\ell0}\delta_{k1}\Bigr],
  \label{eq:lbd_test_lg}\\
  \mathcal{C}^{F\ell k}_{\alpha\beta} &= \nu_{\alpha\beta}\biggl[
      \lambda_{\alpha\beta}\sqrt{\frac{m_\alpha T_{\beta0}}{m_\beta T_{\alpha0}}}\,
      M^{10}_\beta\,\delta_{\ell1}\delta_{k0}
      \nonumber\\
    &\qquad\quad
      + \Bigl(2\kappa_{\alpha\beta}\,\frac{T_{\beta0}}{T_{\alpha0}}\,M^{01}_\beta
      - \sqrt6\,(\theta_{\alpha\beta}-1)\,M^{00}_\beta\Bigr)\,\delta_{\ell0}\delta_{k1}\biggr],
  \label{eq:lbd_final_lg}
\end{align}
with $g_{\ell k}=-2\sqrt{k(k+\ell+\tfrac12)}$. We remark that, as on the HL basis,
the self-restoring and field terms act only on the momentum row $(1,0)$ and on the
energy row $(0,1)$, the latter being, on the LG basis, a single moment. Indeed,
Eqs.~\eqref{eq:lbd_test_lg} and \eqref{eq:lbd_final_lg} follow also from
Eqs.~\eqref{eq:lbd_test_hl} and \eqref{eq:lbd_field_hl} by using
$\Psi_{01}=(\sqrt2\,\Phi_{01}-\Phi_{20})/\sqrt3$.

Two inter-species frequencies are available for the
multi-species LBD operator \cite{ulbl2022,francisquez2022}. They differ by a mass factor multiplying $\nu_{\alpha\beta}$ of Eq.~\eqref{eq:nu_ab},
\begin{equation}
  \nu^{\EM}_{\alpha\beta} = \nu_{\alpha\beta}\,
    \frac{\hat m_\alpha+\hat m_\beta}{2\hat m_\alpha}
  \qquad\text{and}\qquad
  \nu^{\ET}_{\alpha\beta} = \nu_{\alpha\beta},
  \label{eq:nu_lbd_emet}
\end{equation}
Eq.~\eqref{eq:nu_ab} being itself the $\ET$ form. The $\ET$ form is symmetric,
$n_\alpha\nu_{\alpha\beta}=n_\beta\nu_{\beta\alpha}$, such that
Eq.~\eqref{eq:lbd_weights} yields $\lambda_{\alpha\beta}=m_\beta/(m_\alpha+m_\beta)$ and
$\kappa_{\alpha\beta}=\tfrac12$, i.e., the mixing flow is the center-of-mass flow. On
the other hand, the $\EM$ form satisfies detailed balance,
$m_\alpha n_\alpha \nu_{\alpha\beta} = m_\beta n_\beta \nu_{\beta\alpha}$, such that
$\lambda_{\alpha\beta}=\tfrac12$ and $\kappa_{\alpha\beta}=m_\alpha/(m_\alpha+m_\beta)$.

\section{NC Diagnostics}
\label{app:diag}

We report here the NC diagnostics and their normalization. The radial fluxes of particles and kinetic energy follow as velocity-space integrals of $f_{\alpha1}$ weighted by the radial magnetic drift and, together with the bootstrap current, are given by
\begin{align}
  \Gamma_\alpha &= \Bigl\langle \int d^3v\;
     \bm{V}_{D\alpha}\cdot\nabla\psi\; f_{\alpha1}\Bigr\rangle,
  \nonumber\\
  Q_\alpha &= \Bigl\langle \int d^3v\;\frac{m_\alpha v^{2}}{2}\,
     \bm{V}_{D\alpha}\cdot\nabla\psi\; f_{\alpha1}\Bigr\rangle,
  \nonumber\\
  \jboot &= \sum_\alpha q_\alpha n_\alpha
     \bigl\langle \hat B u_{\parallel\alpha}\bigr\rangle,
  \label{eq:fluxdefs}
\end{align}
with $n_\alpha u_{\parallel\alpha} = \int d^3v\; v_\parallel f_{\alpha 1}$ and $\hat B = B/B_\mathrm{ref}$ the normalized field
amplitude. The angle brackets in Eq.~\eqref{eq:fluxdefs} denote the flux-surface average operator.

Using the definitions of the curvature and $\nabla B$ components of the radial magnetic
drift introduced below Eq.~\eqref{eq:dke_cf}, the NC quantities of
Eq.~\eqref{eq:fluxdefs} can be expressed as velocity-space moments of $f_{\alpha 1}$
weighted by the parallel and perpendicular energies, such that
\begin{align}
  \Gamma_\alpha &= V_{D\alpha} \Bigl\langle
       \hat K^{\kappa}_\psi \int d^3v\; 2 c_{\parallel\alpha}^{2}\, f_{\alpha1}
     + \hat K^{\nabla B}_\psi \int d^3v\; c_{\perp\alpha}^{2}\, f_{\alpha1}
     \Bigr\rangle,
  \nonumber\\
  Q_\alpha &= T_\alpha V_{D\alpha} \Bigl\langle
       \hat K^{\kappa}_\psi \int d^3v\; 2 c_{\parallel\alpha}^{2}\, c_\alpha^{2}\,
       f_{\alpha1}
     + \hat K^{\nabla B}_\psi \int d^3v\; c_{\perp\alpha}^{2}\, c_\alpha^{2}\,
       f_{\alpha1} \Bigr\rangle,
  \label{eq:fluxdefs_split}
\end{align}
where we introduce the dimensionless geometric
coefficients $\hat K^{\kappa}_\psi = (R_0 / \hat B) \bm b \times \bm \kappa \cdot
\nabla \psi \left(d \psi / d r\right)^{-1}$ and $\hat K^{\nabla B}_\psi = (R_0 / \hat B)
\bm b \times \nabla \ln B \cdot \nabla \psi \left(d \psi / d r\right)^{-1}$, being
$d \psi / d r$ the radial metric evaluated at the reference surface, such that
$V^{s}_{D\alpha} = (m_\alpha/T_\alpha) V_{D\alpha} \hat K^{s}_\psi$ (for $s = \kappa$,
$\nabla B$), with $V_{D\alpha} = T_\alpha (d\psi/dr) / (q_\alpha B_\mathrm{ref} R_0)$ the
drift amplitude. The velocity-space terms appearing in Eqs.~\eqref{eq:fluxdefs_split} are polynomials of low degree and are therefore represented exactly by a finite number of
components in both bases. More precisely, in the HL basis of Eq.~\eqref{eq:hl_expand}, we have
\begin{align}
  2 c_{\parallel\alpha}^{2} &= \Phi_{00} + \sqrt2\,\Phi_{20},
  \nonumber\\
  c_{\perp\alpha}^{2} &= \Phi_{00} - \Phi_{01},
  \nonumber\\
  2 c_{\parallel\alpha}^{2} c_\alpha^{2} &= \frac52\,\Phi_{00} - \Phi_{01}
     + 4\sqrt2\,\Phi_{20} - \sqrt2\,\Phi_{21} + \sqrt6\,\Phi_{40},
  \nonumber\\
  c_{\perp\alpha}^{2} c_\alpha^{2} &= \frac52\,\Phi_{00} - \frac92\,\Phi_{01}
     + 2\,\Phi_{02} + \frac{1}{\sqrt2}\,\Phi_{20} - \frac{1}{\sqrt2}\,\Phi_{21},
  \label{eq:vdr_hl}
\end{align}
while, in the LG basis of Eq.~\eqref{eq:lg_expand}, the corresponding terms become
\begin{align}
  2 c_\alpha^{2}\xi^{2} &= \Psi_{00}
     - \sqrt{\frac23}\,\Psi_{01}
     + \frac{2}{\sqrt3}\,\Psi_{20},
  \nonumber\\
  c_\alpha^{2}\bigl(1-\xi^{2}\bigr) &= \Psi_{00}
     - \sqrt{\frac23}\,\Psi_{01}
     - \frac{1}{\sqrt3}\,\Psi_{20},
  \nonumber\\
  2 c_\alpha^{4}\xi^{2} &= \frac52\,\Psi_{00}
     - 5\sqrt{\frac23}\,\Psi_{01}
     + \frac{\sqrt{30}}{3}\,\Psi_{02}
     + \frac{7}{\sqrt3}\,\Psi_{20}
     - \frac{\sqrt{42}}{3}\,\Psi_{21},
  \nonumber\\
  c_\alpha^{4}\bigl(1-\xi^{2}\bigr) &= \frac52\,\Psi_{00}
     - 5\sqrt{\frac23}\,\Psi_{01}
     + \frac{\sqrt{30}}{3}\,\Psi_{02}
     - \frac{7}{2\sqrt3}\,\Psi_{20}
     + \frac{\sqrt{42}}{6}\,\Psi_{21}.
  \label{eq:vdr_lg}
\end{align}
Eqs.~\eqref{eq:vdr_hl} and \eqref{eq:vdr_lg} are exact for $P \ge 4$, $J \ge 2$ and $L \ge 2$, $K \ge 2$, respectively. Replacing $f_{\alpha 1}$ by its spectral expansion, closed forms of Eq.~\eqref{eq:fluxdefs} in terms of the spectral moments follow from the recursive properties of the polynomials. The NC quantities are normalized in gB units,
i.e.,
\begin{equation}
  \Gamma_\mathrm{gB} = n_\mathrm{ref}c_\mathrm{ref}\rho_*^{2},
  \qquad
  Q_\mathrm{gB} = n_\mathrm{ref}T_\mathrm{ref}c_\mathrm{ref}\rho_*^{2},
  \qquad
  j_\mathrm{gB} = e\,n_\mathrm{ref}c_\mathrm{ref}\rho_*,
  \label{eq:gyrobohm}
\end{equation}
being $\rho_* = \rho_\mathrm{ref}/L_\mathrm{ref}$ and $\rho_\mathrm{ref} =
m_\mathrm{ref}c_\mathrm{ref}/(eB_\mathrm{ref})$.

\bibliography{library}

\end{document}